\documentclass[aps, pre, reprint, amsmath, amssymb, floatfix]{revtex4-2}

\usepackage{graphicx}
\usepackage{bm}

\usepackage{textcomp,marvosym}
\usepackage{float}

\usepackage[normalem]{ulem}
\usepackage{xcolor}
\usepackage{algorithm}
\usepackage{algpseudocode}
\usepackage{stmaryrd} 
\usepackage{comment}

\begin{document}

\title{
Data-driven reconstruction of spatiotemporal phase dynamics for traveling and oscillating patterns via Bayesian inference
}

\author{Takahiro Arai$^1$}
\email{araitak@jamstec.go.jp}

\author{Toshio Aoyagi$^2$}

\author{Yoji Kawamura$^1$}

\affiliation{$^1$Center for Mathematical Science and Advanced Technology, Japan Agency for Marine-Earth Science and Technology, Yokohama 236-0001, Japan}

\affiliation{$^2$Graduate School of Informatics, Kyoto University, Yoshida-Honmachi, Sakyo-ku, Kyoto 606-8501, Japan}

\date{\today}

\clearpage
\begin{abstract}
Building on the phase reduction theory formulated for reaction-diffusion systems with spatial translational symmetry, we develop a data-driven method that reconstructs the spatiotemporal phase dynamics of traveling and oscillating patterns.
Spatiotemporal phase dynamics are described by spatial and temporal phases that represent the position and oscillation of the pattern, respectively.
Using Bayesian inference, our method directly reconstructs phase equations from time-series data.
When tested on simulation data from coupled Gray-Scott models exhibiting traveling breathers, the method accurately reconstructs the deterministic part of the phase equations in the weak-noise regime, in which the phase dynamics converge to a linearly stable fixed point.
\end{abstract}

\maketitle

\clearpage

\section{INTRODUCTION}
\label{sec:intro}

\begin{figure*}[!t]
    \begin{center}
        \includegraphics[scale=0.95]{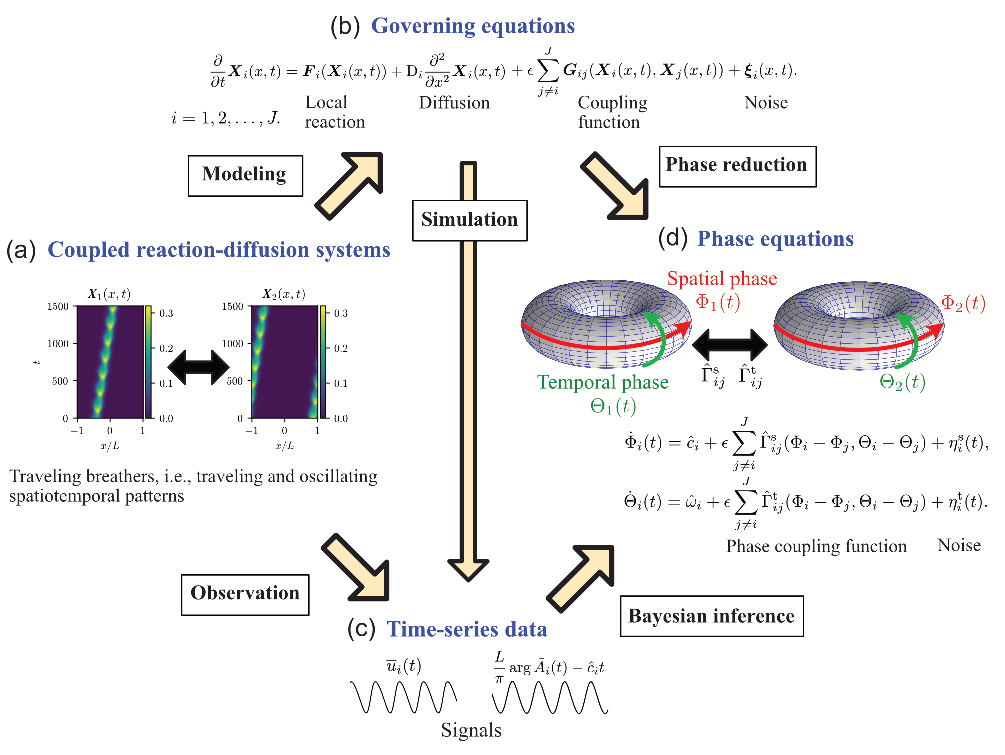}
        \caption{
            Schematic of data-driven and model-driven approaches for phase reduction of weakly coupled limit-torus oscillators.
            (a)~Coupled reaction-diffusion systems. 
            We consider systems exhibiting traveling breathers under medium homogeneity and periodic boundary conditions, which ensure spatial translational symmetry.
            (b)~Governing equations of coupled reaction-diffusion systems.
            (c)~Time-series data.
            To calculate the phase time series, signals are constructed from the spatiotemporal data $\overline{u_i}(t)$ and $\frac{L}{\pi}\arg \tilde{A}_i(t) - \hat{c}_i t$.
            (d)~Phase equations describing the spatiotemporal phase dynamics of the breathers.
            The spatial and temporal phases represent position and oscillation, respectively.
            A model-driven approach follows path (a) $\rightarrow$ (b) $\rightarrow$ (d). 
            The governing equations are formulated through modeling and the phase equations are derived via phase reduction.
            A data-driven approach (developed in this study) follows path (a) $\rightarrow$ (c) $\rightarrow$ (d). 
            The time-series data are obtained through observation and the phase equations are reconstructed directly using Bayesian inference.
            To validate the data-driven approach, we numerically simulate the governing equations and use the resulting time-series data to reconstruct the phase equations.
            This workflow follows path (b) $\rightarrow$ (c) $\rightarrow$ (d).
            To quantify the accuracy of our method, we use the ground truth (the phase equations derived from the governing equations through phase reduction). 
            This workflow follows path (b) $\rightarrow$ (d).
        }
        \label{fig:fig1}
    \end{center}
\end{figure*}

\par
Synchronization phenomena in various domains, including physics, chemistry, and life sciences, are often studied as rhythmic systems described by self-sustained oscillators with stable limit-cycle solutions~\cite{kuramoto_chemical_1984,winfree_geometry_1980,pikovsky_synchronization_2001,manrubia_emergence_2004,osipov_synchronization_2007,kuramoto_half_2026}.
Phase reduction theory is a useful framework for analyzing the synchronization properties of coupled oscillators~\cite{neu_coupled_1979,brown_phase_2004,ermentrout_type_1996,nakao_phase_2016,pietras_network_2019,ermentrout_mathematical_2010,hoppensteadt_weakly_1997,izhikevich_dynamical_2006,kuramoto_concept_2019} and collective dynamics in oscillator networks, including globally coupled, nonlocally coupled, and complex network systems~\cite{acebron_kuramoto_2005,arenas_synchronization_2008,ashwin_mathematical_2016,pikovsky_dynamics_2015,stankovski_coupling_2017,strogatz_kuramoto_2000,rodrigues_kuramoto_2016}.
Phase reduction theory can be viewed as a model-driven approach, providing a foundation for data-driven approaches that reconstruct causal relationships from time-series data in the form of phase descriptions~\cite{stankovski_coupling_2017}.

\par
Data-driven approaches based on the phase reduction theory have enabled the identification of coupling directions~\cite{rosenblum_detecting_2001} and the reconstruction of phase sensitivity functions~\cite{galan_efficient_2005,ermentrout_relating_2007,ota_weighted_2009,imai_robust_2017,cestnik_reconstructing_2017,cestnik_inferring_2018} and phase coupling functions~\cite{stankovski_coupling_2017,miyazaki_determination_2006,tokuda_inferring_2007,kralemann_uncovering_2007,kralemann_phase_2008,stankovski_inference_2012,duggento_dynamical_2012,kralemann_vivo_2013,ota_direct_2014,arai_extracting_2022,yamaguchi_reconstruction_2024,matsuki_network_2025,kato_bayesian_2025,yoneda_gaussian_2025,matsuki_isochron-free_2026}.
Several studies have improved the calculation of phase time series from observations, thereby enabling more accurate reconstruction of phase dynamics~\cite{cestnik_reconstructing_2017,cestnik_inferring_2018,kralemann_uncovering_2007,kralemann_phase_2008,revzen_estimating_2008,namura_estimating_2022,rosenblum_real-time_2021,yamamoto_gaussian_2025,wilshin_estimating_2025,gengel_phase_2022,matsuki_extended_2023}.
Machine learning techniques beyond the phase reduction framework, including Koopman-based methods~\cite{shirasaka_phase-amplitude_2017,mauroy_global_2018}, autoencoders~\cite{fukami_data-driven_2024,yawata_phase_2024}, and recurrent neural network approaches~\cite{cestnik_inferring_2019}, have also been explored.
These developments provide powerful tools for uncovering the mechanisms underlying synchronization phenomena such as locomotor coordination in humans and millipedes~\cite{arai_interlimb_2024,furukawa_bayesian_2025,funato_evaluation_2016,yoshikawa_wavy_nodate}, brain waves~\cite{stankovski_coupling_2017,onojima_dynamical_2018}, the human cardiorespiratory system~\cite{kralemann_vivo_2013}, selective attention mechanisms in frog choruses~\cite{ota_interaction_2020,aihara_gradual_2026}, and acoustic interactions between male cicadas~\cite{ishimaru_temporal_2025}.

\par
Synchronization occurs not only in oscillators but also in spatiotemporal dynamics. 
For example, meteorological studies have reported the synchronization of sea surface temperatures between the Kuroshio Current and the Gulf Stream~\cite{cessi_gulf_2021,kohyama_gulf_2021,kohyama_interactive_2025,yamagami_gulf_2025,yasuda_mechanism_2025,dong_seasonal_2026} and the synchronization between atmospheric variability patterns such as the Arctic and Antarctic Oscillations~\cite{tachibana_interhemispheric_2018}. 
In chemistry, synchronization between a pair of photosensitive Belousov-Zhabotinsky systems coupled via video cameras and projectors was experimentally demonstrated~\cite{hildebrand_synchronization_2003}.

\par 
Model-driven approaches to analyzing interactions between spatiotemporal patterns include finite-dimensional reduction methods that are distinct from the phase reduction framework considered here.
For example, in reaction-diffusion systems, the dynamics of interacting pulses~\cite{ohta_pulse_2001,ei_motion_2002,ei_pulsepulse_2002,nishiura_arbitrarily_2023}, spots~\cite{ei_interacting_2006}, and fronts~\cite{nishi_reduction_2019} have been described using reduced equations for collective coordinates such as position, velocity, and oscillation amplitude.
Reduced equations have also been used to control the motion of traveling waves~\cite{lober_controlling_2014}.
Related studies of interacting breathers have examined the selection between in-phase and out-of-phase oscillations~\cite{ikeda_pattern_1994}, as well as the phase dependence of collision dynamics~\cite{teramoto_phase-dependent_2004}.
The analysis of synchronization phenomena in spatiotemporal dynamics has also led to the development of phase reduction theory for partial differential equations with stable limit-cycle solutions.
The theory has been formulated for a wide range of systems, including oscillatory convection~\cite{kawamura_collective_2013,kawamura_noise-induced_2014}, reaction-diffusion systems~\cite{nakao_phase-reduction_2014,kawamura_optimizing_2017,ozawa_phase_2026}, periodic flows~\cite{taira_phase-response_2018,iima_jacobian-free_2019,iima_phase_2021,kawamura_adjoint-based_2022,godavarthi_optimal_2023,iima_optimal_2024}, and beating flagella~\cite{kawamura_phase_2018}.
In these frameworks, the dynamics of a spatiotemporal system are reduced to a single phase variable.

\par
Recently, data-driven approaches have also been proposed to reconstruct spatiotemporal phase dynamics from observed data.
These include approaches based on autoencoders~\cite{fukami_data-driven_2024,yawata_phase_2025}, as well as a method we previously proposed for calculating phase time series using Poincar\'e sections~\cite{arai_setting_2025}.
However, data-driven approaches for quantitatively analyzing synchronization phenomena in traveling and oscillating spatiotemporal patterns remain underdeveloped.
Such patterns include breathers in reaction-diffusion systems~\cite{yadome_chaotic_2011,hagberg_pattern_1994} and oscillatory convection in fluid systems~\cite{kawamura_phase_2015,kawamura_phase_2019}.
Phase reduction theory for partial differential equations with spatial translational symmetry describes the dynamics of such patterns in terms of spatial and temporal phases, representing their position and oscillation, respectively~\cite{arai_phase_2025,kawamura_phase_2015,kawamura_phase_2019}.
The temporal phase is analogous to that in conventional phase reduction theory for limit-cycle solutions, whereas the spatial phase arises from spatial translational symmetry.
The theory also shows that the spatial and temporal phases are mutually coupled.
Therefore, analyzing synchronization properties from observed data requires data-driven approaches that quantitatively reconstruct phase dynamics while incorporating the effects of the spatial phase, i.e., the translation of the spatiotemporal pattern.


\par 
On the basis of the phase reduction framework for traveling breathers in reaction-diffusion systems, we develop a Bayesian inference method~\cite{bishop_pattern_2006} that reconstructs the spatiotemporal phase dynamics.
We test our approach on simulation data of coupled Gray-Scott models exhibiting traveling breathers~\cite{arai_phase_2025,yadome_chaotic_2011}.
Because the governing equations are known, phase equations derived through a model-driven approach (namely, phase reduction analysis) provide the ground truth for assessing the accuracy of the proposed method~\cite{arai_phase_2025}.


\par
Figure~\ref{fig:fig1} provides an overview of the present study, illustrating model-driven and data-driven approaches to phase reduction of weakly coupled limit-torus oscillators. 
A pair of traveling breathers in reaction-diffusion systems is considered (Fig.~\ref{fig:fig1}(a)), where spatial translational symmetry arises from medium homogeneity and periodic boundary conditions.
In the model-driven approach, governing equations are formulated as a mathematical model of the systems (Fig.~\ref{fig:fig1}(b)), whereas in the data-driven approach, time-series data are obtained from observations of spatiotemporal dynamics or from numerical simulations (Fig.~\ref{fig:fig1}(c)). 
Phase equations for the spatial and temporal phases, which represent the position and oscillation of the breathers, respectively, are then obtained either by phase reduction of the governing equations or by reconstruction from time-series data using Bayesian inference (Fig.~\ref{fig:fig1}(d)). 
Accordingly, the model-driven approach follows the path (a) $\to$ (b) $\to$ (d), whereas the data-driven approach follows the path (a) $\to$ (c) $\to$ (d).
In this study, we develop a data-driven approach based on the phase reduction theory underlying the model-driven approach, which has been established in Refs.~\cite{arai_phase_2025,kawamura_phase_2015,kawamura_phase_2019}.
To validate the proposed method, we examine whether phase equations can be accurately reconstructed from simulation data of coupled Gray-Scott models; this follows the path (b) $\to$ (c) $\to$ (d).
The reconstruction accuracy is then evaluated by comparison with phase equations analytically derived from the governing equations via phase reduction, which serve as the ground truth; this corresponds to the path (b) $\to$ (d).

\par
This paper is organized as follows.
In Sec.~\ref{sec:sec2}, we present the data-driven approach for estimating phase equations of traveling breathers.
In Sec.~\ref{sec:sec3}, we test the proposed method using simulation data of coupled Gray-Scott models and verify its performance.
Finally, concluding remarks are given in Sec.~\ref{sec:concluding_remarks}.
Supplementary information is provided in Appendixes~\ref{appendix:noise_derivation}, \ref{appendix:Hilbert}, \ref{appendix:IndependentInference}, and \ref{appendix:SpatiallyUniformNoise}.

\section{Data-driven Approach for Reconstructing Spatiotemporal Phase Dynamics}
\label{sec:sec2}

\par
In this section, we present a data-driven approach for reconstructing the phase equations of traveling and oscillating spatiotemporal dynamics.
We first introduce the problem setting based on phase description~\cite{arai_phase_2025} (Sec.~\ref{subsec:ReviewPhaseReduction}), then describe how the spatial and temporal phases are extracted from spatiotemporal data (Sec.~\ref{subsec:CalculationOfPhase}), and finally present a Bayesian inference method for reconstructing the phase equations (Sec.~\ref{subsec:BayesianInference}).
Among these subsections, Secs.~\ref{subsec:CalculationOfPhase} and \ref{subsec:BayesianInference} present the data-driven approach proposed in this study.

\subsection{Problem setting based on phase description}
\label{subsec:ReviewPhaseReduction}

\par
We consider the following one-dimensional reaction-diffusion system:
\begin{align}
    \label{eq:general_PDE}
    \frac{\partial}{\partial t} \boldsymbol{X}(x, t)
    =  \boldsymbol{F}(\boldsymbol{X}(x,t)) 
    + {\rm D} \frac{\partial^2}{\partial x^2} \boldsymbol{X}(x,t),
\end{align}
where $\boldsymbol{X}(x, t)$ denotes the state of the medium at point $x$ at time $t$, $\boldsymbol{F}(\boldsymbol{X}(x,t))$ denotes the local reaction dynamics, and ${\rm D} \frac{\partial^2}{\partial x^2} \boldsymbol{X}(x,t)$ denotes the diffusion term of $\boldsymbol{X}(x,t)$ over the medium, with ${\rm D}$ being the diffusion coefficient matrix.
The system size is $2L$, i.e., $x \in [0, 2L)$. 
We impose $2L$-periodic boundary conditions as follows:
\begin{align}
    \label{eq:periodic_condition}
    \boldsymbol{X}(x + 2L, t) = \boldsymbol{X}(x, t).
\end{align}
Owing to the homogeneous medium in Eq.~(\ref{eq:general_PDE}) and the periodic boundary condition on $x$, this system exhibits continuous spatial translational symmetry with respect to $x$.

\par
We assume that the reaction-diffusion system described in Eq.~(\ref{eq:general_PDE}) has a stable limit-torus solution, representing a traveling and oscillating spatiotemporal pattern, i.e., a traveling breather.
The limit-torus solution is described as follows:
\begin{align}
    \label{eq:limit-torus}
    & \boldsymbol{X}(x,t) = \boldsymbol{X}_0(x - \Phi(t), \Theta(t)), ~
    \dot{\Phi}(t) = c, ~
    \dot{\Theta}(t) = \omega,
\end{align}
where $c$ and $\omega$ are the traveling velocity and oscillation frequency of the breather, respectively.
The spatial phase $\Phi \in [0, 2L)$ and temporal phase $\Theta \in [0, 2\pi)$ represent the position and oscillation of the spatiotemporal pattern, respectively.
The traveling velocity $c$ and oscillation frequency $\omega$ are constant.
A breather with $c \neq 0$ is called a traveling breather~\cite{yadome_chaotic_2011}.

\par 
Next, we consider weakly coupled reaction-diffusion systems with noise, which are described by the following equation ($i=1,2,\ldots, J$):
\begin{align}
\label{eq:PDE_CoupledModel}
\frac{\partial}{\partial t} \boldsymbol{X}_i(x, t)
=  & \boldsymbol{F}_i(\boldsymbol{X}_i(x,t))
+ {\rm D}_i \frac{\partial^2}{\partial x^2} \boldsymbol{X}_i(x,t) \nonumber \\
& + \epsilon \sum_{j \neq i}^J \boldsymbol{G}_{ij}(\boldsymbol{X}_i(x,t), \boldsymbol{X}_j(x,t))
+ \boldsymbol{\xi}_i(x,t),
\end{align}
where subscripts $i$ and $j$ denote indices of the systems, and $\epsilon \boldsymbol{G}_{ij}(\boldsymbol{X}_i(x,t), \boldsymbol{X}_j(x,t))$ represents the coupling function between the $i$-th and $j$-th systems at a spatial point $x$.
We assume sufficiently weak coupling ($\epsilon \ll 1$) and sufficiently small noise, so that the solution $\boldsymbol{X}_i(x,t)$ remains in the vicinity of the orbit of the limit-torus solution, $\boldsymbol{X}_0(x-\Phi_i,\Theta_i)$.
We further assume that the noise term $\boldsymbol{\xi}_i(x,t)$ is Gaussian spatial block noise.
In each spatial interval $x \in [\alpha \Delta, (\alpha+1)\Delta)$ with $\Delta = 2L/K$, the noise is given by
\begin{align}
    \label{eq:blocknoise_xi}
    \boldsymbol{\xi}_i(x,t) = \boldsymbol{\xi}_{i,\alpha}(t).
\end{align}
For $\alpha, \beta = 0, 1, \ldots, K-1$, the statistics of the spatial block noise are given by
\begin{align}
    \label{eq:mean_xi}
    & \langle \boldsymbol{\xi}_{i, \alpha}(t) \rangle 
    = \boldsymbol{0}, \\
    \label{eq:cov_xi}
    & \langle \boldsymbol{\xi}_{i, \alpha}(t)\, [\boldsymbol{\xi}_{j, \beta}(s)]^\mathrm{T} \rangle
    = \sigma_i^2\, {\rm I}\, \delta_{ij}\,\delta_{\alpha\beta}\,  \,\delta(t - s), 
\end{align}
where $\delta(t - s)$ and $\delta_{ij}$ denote the Dirac delta function and the Kronecker delta, respectively.
For simplicity, we assume that the noise intensity $\sigma_i^2$ is identical for all spatial blocks and components.

\par
The representation of spatial block noise corresponds directly to the spatial discretization used in the numerical simulations of the reaction-diffusion models.
For this type of noise, a formulation describing how stochastic perturbations are projected onto reduced variables has been discussed~\cite{masuda_collective_2010}.
Applying this formulation to the phase reduction theory for limit-torus solutions~\cite{arai_phase_2025}, we derive the effective noise intensity in the phase equations (see Appendix~\ref{appendix:noise_derivation}).


\par
On the basis of the phase reduction analysis~\cite{arai_phase_2025, kawamura_phase_2015, kawamura_phase_2019}, the following phase equations for the spatial and temporal phases are derived~\cite{note:noise_induced_shifts}:
\begin{align}
    \label{eq:PhaseEquation_Phi}
    \dot{\Phi}_i(t) 
    &= c_i + \epsilon \sum_{j \neq i}^J \Gamma_{ij}^{\mathrm{s}}(\Phi_i - \Phi_j, \Theta_i - \Theta_j) + \eta_{i}^{\mathrm{s}}(t), \\
    \label{eq:PhaseEquation_Theta}
    \dot{\Theta}_i(t)
    &= \omega_i + \epsilon \sum_{j \neq i}^J \Gamma_{ij}^{\mathrm{t}}(\Phi_i - \Phi_j, \Theta_i - \Theta_j) + \eta_{i}^{\mathrm{t}}(t).
\end{align}
The phase coupling functions, $\Gamma_{ij}^{\mathrm{s}}(\Phi_i - \Phi_j, \Theta_i - \Theta_j)$ and $\Gamma_{ij}^{\mathrm{t}}(\Phi_i - \Phi_j, \Theta_i - \Theta_j)$, depend only on the spatial and temporal phase differences. 
The noise terms are denoted by $\eta_i^{\mathrm{s}}(t)$ and $\eta_i^{\mathrm{t}}(t)$.
The phase coupling functions are expanded in a two-dimensional Fourier series as
\begin{align}
    \label{eq:Gamma_s_expansion}
    & \epsilon \Gamma_{ij}^{\mathrm{s}}(\Phi_i - \Phi_j, \Theta_i - \Theta_j)  \nonumber \\
    & = a_{ij,\boldsymbol{0}}^{\mathrm{s}}
    + \sum_{\boldsymbol{m} \in \mathbb{M}} a_{ij,\boldsymbol{m}}^{\mathrm{s}}\,
      g_{\boldsymbol{m}} (\Phi_i - \Phi_j, \Theta_i - \Theta_j),
    \\
    \label{eq:Gamma_t_expansion}
    & \epsilon \Gamma_{ij}^{\mathrm{t}}(\Phi_i - \Phi_j, \Theta_i - \Theta_j)  \nonumber \\
    & = a_{ij,\boldsymbol{0}}^{\mathrm{t}}
    + \sum_{\boldsymbol{m} \in \mathbb{M}} a_{ij,\boldsymbol{m}}^{\mathrm{t}}\,
      g_{\boldsymbol{m}} (\Phi_i - \Phi_j, \Theta_i - \Theta_j),
\end{align}
where $a_{ij,\boldsymbol{m}}^{\mathrm{s}}$ and $a_{ij,\boldsymbol{m}}^{\mathrm{t}}$ are the Fourier coefficients, and $a_{ij,\boldsymbol{0}}^{\mathrm{s}}$ and $a_{ij,\boldsymbol{0}}^{\mathrm{t}}$ denote the constant parts of the phase coupling functions.
The index vector $\boldsymbol m=(m_{\mathrm s},m_{\mathrm t}) \in \mathbb{Z}^2$ is taken from the set
\begin{align}
    \mathbb{M}
    := \left\{
        \boldsymbol m \;\middle|\;
        |m_{\mathrm s}|\le M_{\mathrm s},\,
        |m_{\mathrm t}|\le M_{\mathrm t},\,
        \boldsymbol m\neq \boldsymbol 0
    \right\}, 
\end{align}
where $m_{\mathrm s}$ and $m_{\mathrm t}$ denote the Fourier modes associated with the spatial and temporal phase differences, respectively.
The complex Fourier basis functions are defined by
\begin{align}
    \label{eq:basisfunction_g}
    & g_{\boldsymbol{m}} (\Phi_i - \Phi_j, \Theta_i - \Theta_j )
    \nonumber \\
    & := \exp \left[
        \mathrm{i} \left(
            m_\mathrm{s} \frac{\pi(\Phi_i - \Phi_j)}{L}
            + m_\mathrm{t} (\Theta_i - \Theta_j)
        \right)
    \right].
\end{align}

\par
To eliminate redundancy in the constant terms, we define the following constants: 
    \begin{align}
        \label{eq:define_c_hat}
        & \hat{c}_i := c_i + \sum_{j \neq i}^J a_{ij,\boldsymbol{0}}^{\mathrm{s}}  + \Delta c_i,  \\
        \label{eq:define_omega_hat}
        & \hat{\omega}_i := \omega_i + \sum_{j \neq i}^J a_{ij,\boldsymbol{0}}^{\mathrm{t}} + \Delta \omega_i,
    \end{align}
and the following functions:
    \begin{align}
        \label{eq:define_Gamma_s_hat}
        & \epsilon \hat{\Gamma}_{ij}^{\mathrm{s}}(\Phi_i - \Phi_j, \Theta_i - \Theta_j) \nonumber \\
        & := \sum_{\boldsymbol{m} \in \mathbb{M}} a_{ij,\boldsymbol{m}}^{\mathrm{s}}\, g_{\boldsymbol{m}} (\Phi_i - \Phi_j, \Theta_i - \Theta_j), \\
        \label{eq:define_Gamma_t_hat}
        &  \epsilon \hat{\Gamma}_{ij}^{\mathrm{t}}(\Phi_i - \Phi_j, \Theta_i - \Theta_j)  \nonumber \\
        & := \sum_{\boldsymbol{m} \in \mathbb{M}} a_{ij,\boldsymbol{m}}^{\mathrm{t}}\, g_{\boldsymbol{m}} (\Phi_i - \Phi_j, \Theta_i - \Theta_j).
    \end{align}
In Eqs.~(\ref{eq:define_c_hat})--(\ref{eq:define_Gamma_t_hat}), the constant parts of the phase coupling functions for the spatial and temporal phases, i.e.,
$\sum_{j \neq i}^{J} a_{ij,\boldsymbol{0}}^{\mathrm{s}}$ and
$\sum_{j \neq i}^{J} a_{ij,\boldsymbol{0}}^{\mathrm{t}}$,
are absorbed into $\hat{c}_i$ and $\hat{\omega}_i$, respectively.
Hereafter, we refer to
$\hat{\Gamma}_{ij}^{\mathrm{s}}(\Phi_i-\Phi_j,\Theta_i-\Theta_j)$ and
$\hat{\Gamma}_{ij}^{\mathrm{t}}(\Phi_i-\Phi_j,\Theta_i-\Theta_j)$
simply as the phase coupling functions.
Furthermore, the noise-induced shifts $\Delta c_i$ and $\Delta \omega_i$, which depend on $\sigma_i^2$, are explicitly incorporated into $\hat{c}_i$ and $\hat{\omega}_i$, respectively.

\par
In the Bayesian inference described in Sec.~\ref{subsec:BayesianInference},
$\hat{c}_i$ is estimated directly, without separately determining
$c_i$, $\{a_{ij,\boldsymbol{0}}^{\mathrm{s}}\}_{j\ne i}$, and $\Delta c_i$.
Similarly, $\hat{\omega}_i$ is estimated directly, without separately determining $\omega_i$, $\{a_{ij,\boldsymbol{0}}^{\mathrm{t}}\}_{j\ne i}$, and $\Delta\omega_i$.
Theoretical expressions for noise-induced shifts have been derived for limit-cycle oscillators described by ordinary differential equations~\cite{yoshimura_phase_2008,teramae_stochastic_2009,goldobin_dynamics_2010,nakao_effective_2010}, but not for limit-torus oscillators described by partial differential equations.
Therefore, we define reference values for the constant terms based on the traveling velocities and oscillation frequencies measured from numerical simulations of the uncoupled systems with noise (Sec.~\ref{subsec:Results_of_BayesianInference}).
These reference values are treated as the true values when comparing them with the estimated constant terms.

\par 
Using Eqs.~(\ref{eq:define_c_hat})--(\ref{eq:define_Gamma_t_hat}), we rewrite the phase equations as
\begin{align}
    \label{eq:PhaseEquation_Phi_hat}
    \dot{\Phi}_i(t)
    &= \hat{c}_i + \epsilon \sum_{j \neq i}^J \hat{\Gamma}_{ij}^{\mathrm{s}}(\Phi_i - \Phi_j, \Theta_i - \Theta_j) + \eta_{i}^{\mathrm{s}}(t), \\
    \label{eq:PhaseEquation_Theta_hat}
    \dot{\Theta}_i(t)
    &= \hat{\omega}_i + \epsilon \sum_{j \neq i}^J \hat{\Gamma}_{ij}^{\mathrm{t}}(\Phi_i - \Phi_j, \Theta_i - \Theta_j) + \eta_{i}^{\mathrm{t}}(t).
\end{align}
The statistics of the noise vector $\boldsymbol{\eta}_i(t) = (\eta_i^{\mathrm{s}}(t), \eta_i^{\mathrm{t}}(t))^{\mathrm{T}}$ are given by
\begin{align}
    & \langle \boldsymbol{\eta}_i(t) \rangle = \boldsymbol{0},  \\
    & \langle \boldsymbol{\eta}_i(t) \left[ \boldsymbol{\eta}_j(s) \right]^{\mathrm{T}} \rangle
    = {\rm E}_i \, \delta_{ij} \, \delta(t-s).
\end{align}
The components of the covariance matrix are written as
\begin{align}
    {\rm E}_i =
    \begin{pmatrix}
    E_i^{\mathrm{s}\mathrm{s}} & E_i^{\mathrm{s}\mathrm{t}} \\
    E_i^{\mathrm{t}\mathrm{s}} & E_i^{\mathrm{t}\mathrm{t}}
    \end{pmatrix}.
\end{align}
The matrix ${\rm E}_i$ is symmetric, i.e., $E_i^{\mathrm{s}\mathrm{t}} = E_i^{\mathrm{t}\mathrm{s}}$.
Note that the noise-induced shifts are absorbed into the constant terms defined in Eqs.~(\ref{eq:define_c_hat}) and (\ref{eq:define_omega_hat}), while the stochastic fluctuations are represented by the zero-mean noise vector $\boldsymbol{\eta}_i(t)$ with covariance matrix ${\rm E}_i$.
As described in Appendix~\ref{appendix:noise_derivation}, the effective noise in the averaged phase equations, Eqs.~(\ref{eq:PhaseEquation_Phi_hat}) and (\ref{eq:PhaseEquation_Theta_hat}), is additive; that is, the noise terms do not depend on the spatial or temporal phase values.
Appendix~\ref{appendix:noise_derivation} also provides the derivation of the covariance matrix ${\rm E}_i$.

\par 
Finally, the phase equations can be written in Fourier series form as
\begin{align}
    \label{eq:PhiEquation_expansion}
    & \dot{\Phi}_i (t) 
    =  a_{i,\boldsymbol{0}}^{\mathrm{s}}
    + \sum_{j \neq i}^J \sum_{\boldsymbol{m} \in \mathbb{M}} a_{ij,\boldsymbol{m}}^{\mathrm{s}}\, g_{\boldsymbol{m}}(\Phi_i - \Phi_j, \Theta_i - \Theta_j)
    + \eta_{i}^{\mathrm{s}}(t),
    \\
    \label{eq:ThetaEquation_expansion}
    & \dot{\Theta}_i (t) 
    = a_{i,\boldsymbol{0}}^{\mathrm{t}}
    + \sum_{j \neq i}^J \sum_{\boldsymbol{m} \in \mathbb{M}} a_{ij,\boldsymbol{m}}^{\mathrm{t}}\, g_{\boldsymbol{m}}(\Phi_i - \Phi_j, \Theta_i - \Theta_j) 
    + \eta_{i}^{\mathrm{t}}(t), 
\end{align}
where $a_{i,\boldsymbol{0}}^{\mathrm{s}} = \hat{c}_i$ and $a_{i,\boldsymbol{0}}^{\mathrm{t}} = \hat{\omega}_i$.

\subsection{Calculation of the spatial and temporal phases from time-series data}
\label{subsec:CalculationOfPhase}

\par 
We calculate the time series of the spatial and temporal phases from time-series data.
In this calculation, we use one component of $\boldsymbol{X}_0(x-\Phi_i,\Theta_i)$ and $\boldsymbol{X}_i(x,t)$, and denote them by $u_0(x-\Phi_i,\Theta_i)$ and $u_i(x,t)$, respectively.
A time series of $u_i(x,t)$ is obtained by sampling at discrete times $t_n$ with sampling interval $\Delta t$, where $t_n$ denotes the time of the $n$th data point.

\par 
Using the unperturbed solution, we define a periodic function of the temporal phase as follows:
\begin{align}
    \label{eq:spatial_integration_u0}
    & \overline{u_0}(\Theta_i)
    = \int_0^{2L} \mathrm{d}x\, u_0(x-\Phi_i, \Theta_i).
\end{align}
For weak perturbations, $\overline{u_0}(\Theta_i(t))$ is approximated by 
\begin{align}
    \label{eq:spatial_integration_u}
    & \overline{u_i}(t)
    = \int_0^{2L} \mathrm{d}x\, u_i(x, t).
\end{align}
The temporal phase is calculated from $\overline{u_i}(t)$ through linear interpolation as follows:
\begin{align}
    \label{eq:calculation_Theta}
    \Theta_i(t) = 
    2\pi \frac{t - t_{i,k}}{t_{i,k+1} - t_{i,k}}, \quad (t_{i,k} \leq t < t_{i,k+1}),
\end{align}
where $t_{i,k}$ denotes the $k$th time at which $\overline{u_i}(t)$ intersects the Poincar\'e section.

\par
Using the unperturbed solution, we introduce the following complex-valued function of the spatial and temporal phases:
\begin{align}
    \label{eq:A(Phi,Theta)}
    & A \left( \Phi_i, \Theta_i \right) 
    := \int_0^{2L} \mathrm{d}x\, u_0(x - \Phi_i, \Theta_i) \exp\left[ \mathrm{i} \pi \frac{x}{L} \right].
\end{align}
We obtain the following expression from Eq.~(\ref{eq:A(Phi,Theta)}) due to the spatial translational symmetry:
\begin{align}
    \label{eq:arg_A(Phi,Theta)}
    \arg A(\Phi_i, \Theta_i)
    = \frac{\pi}{L} \left( \Phi_i + B(\Theta_i) \right),
\end{align}
where $B(\Theta_i)$ is a $2\pi$-periodic function.
Let $B(\Theta_i)$ satisfy the following condition:
\begin{align}
    \label{eq:B_condition}
    \frac{1}{2\pi} \int_0^{2\pi} \mathrm{d}\Theta_i\, B(\Theta_i)  = 0.
\end{align}
For subsequent calculations of the spatial phase, we need to determine the function $B(\Theta_i)$.

\par
We approximate $A(\Phi_i(t), \Theta_i(t))$ as
\begin{align}
    \label{eq:tilde_A}
    \tilde{A}_i(t)
    := \int_0^{2L} \mathrm{d}x\, u_i(x, t) \exp\left[ \mathrm{i} \pi \frac{x}{L} \right],
\end{align}
and introduce a slow variable for the spatial phase as $\Phi_i(t) := \hat{c}_i t + \varphi_i(t)$. 
Substituting the slow variable into Eq.~(\ref{eq:arg_A(Phi,Theta)}) yields
\begin{align}
    \label{eq:increment_argA}
    \frac{L}{\pi} \arg \tilde{A}_i(t) - \hat{c}_i t 
    = \varphi_i(t) + \tilde{B}_i(\Theta_i(t)),
\end{align}
where $\tilde{B}_i(\Theta_i(t))$ denotes the function $B(\Theta_i)$ determined from the time series of $\tilde{A}_i(t)$.

\par
Equation~(\ref{eq:increment_argA}) provides the basis for calculating the spatial phase.
Because $\tilde{A}_i(t)$ is directly obtained from Eq.~(\ref{eq:tilde_A}), the remaining procedure consists of the following four steps:
\begin{enumerate}
    \item Determine the traveling velocity $\hat{c}_i$.
    \item Approximate the slow variable for the spatial phase, $\varphi_i(t)$, by treating it as constant over each period of the temporal phase.
    \item Determine the waveform of $\tilde{B}_i(\Theta_i)$ using $\hat{c}_i$ and the approximation of $\varphi_i(t)$.
    \item Calculate the time series of the spatial phase $\Phi_i(t)$ using $\tilde{B}_i(\Theta_i(t))$ and $\tilde{A}_i(t)$.
\end{enumerate}
Each step is described in detail in the following paragraphs.

\par 
(Step~1)~
Substituting $t=t_{i,k}$ and $t_{i, k-1}$ into Eq.~(\ref{eq:increment_argA}) and taking the difference between the two resulting expressions yields
    \begin{align}
        \label{eq:Regression_hatc}
        & \frac{L}{\pi} \left[ \arg \tilde{A}_i(t_{i,k}) - \arg \tilde{A}_i(t_{i,k-1}) \right] = \hat{c}_i (t_{i,k} - t_{i,k-1}).
    \end{align}
In deriving Eq.~(\ref{eq:Regression_hatc}), we assume that the slow variable varies negligibly over one period of the temporal phase, i.e., $|\varphi_i(t_{i,k}) - \varphi_i(t_{i,k-1})| \ll 1$.
According to Eq.~(\ref{eq:Regression_hatc}), the traveling velocity $\hat{c}_i$ is obtained from the increment of $\arg \tilde{A}_i(t)$ over a period of the temporal phase, namely from $t_{i,k-1}$ to $t_{i,k}$, where $\Theta_i(t_{i,k-1}) = \Theta_i(t_{i,k}) = 0$.
Given a dataset $\{\tilde{A}_i(t_{i,k})\}_k$, which can be assembled from multiple time series of $\tilde{A}_i(t)$, $\hat{c}_i$ is determined by linear regression based on Eq.~(\ref{eq:Regression_hatc}).

\par 
(Step~2)~
We calculate the slow variable $\varphi_i(t)$.
Integrating Eq.~(\ref{eq:increment_argA}) over a period of the temporal phase, we obtain $\varphi_i(t)$ for $t_{i,k-1} \leq t < t_{i,k}$:
\begin{align}
    \label{eq:calculation_varphi}
    \varphi_i(t) = \frac{1}{t_{i,k}-t_{i,k-1}} \int_{t_{i,k-1}}^{t_{i,k}} \mathrm{d}\tau\,
    \left[ 
        \frac{L}{\pi} \arg \tilde{A}_i(\tau) - \hat{c}_i \tau
    \right],
\end{align}
where $\varphi_i(t)$ is treated as constant over the interval $t_{i,k-1} \leq t < t_{i,k}$.
In deriving Eq.~(\ref{eq:calculation_varphi}), we use 
\begin{align}
    \label{eq:Bhat_condition}
    \frac{1}{ t_{i,k}-t_{i,k-1}} 
    \int_{t_{i,k-1}}^{t_{i,k}} \mathrm{d}\tau\, \tilde{B}_i(\Theta_i(\tau)) = 0,
\end{align}
which follows from Eq.~(\ref{eq:B_condition}).

\par 
(Step~3)~
To determine the function $\tilde{B}_i(\Theta_i)$, we solve a regression problem based on Eq.~(\ref{eq:increment_argA}).
First, we expand $\tilde{B}_i(\Theta_i)$ in a Fourier series as follows:
\begin{align}
    \label{eq:PeriodicFunction_B}
    \tilde{B}_i(\Theta_i) := \sum_{m=1}^{M} \left[ a_{i,2m} \cos(m \Theta_i) + a_{i,2m+1} \sin(m \Theta_i) \right],
\end{align}
where the constant term is zero according to Eq.~(\ref{eq:Bhat_condition}).
Given a dataset $\{\Theta_i(t_n), b_i(t_n)\}_n$, which can be assembled from multiple time series, we determine the Fourier coefficients $\{a_{i,2m},\, a_{i,2m+1}\}_m$ by solving the following regression problem:
\begin{align}
    \label{eq:regression_B}
    & b_i(t_n) = \sum_{m=1}^{M} \left[ a_{i,2m} \cos(m \Theta_i(t_n)) + a_{i,2m+1} \sin(m \Theta_i(t_n)) \right], \\
    \label{eq:data_bi(t)}
    & b_i(t) :=  \frac{L}{\pi} \arg \tilde{A}_i(t) - \hat{c}_i t - \varphi_i(t).
\end{align}

\par 
(Step~4)~
Setting $\Phi_i(t) = \hat{c}_i t + \varphi_i(t)$, Eq.~(\ref{eq:increment_argA}) can be expressed as
\begin{align}
    \label{eq:calculation_Phi}
    \Phi_i(t) = \frac{L}{\pi} \arg \tilde{A}_i(t) - \tilde{B}_i(\Theta_i(t)).
\end{align} 
In this expression, the waveform of $\tilde{B}_i(\Theta_i)$ is determined by the regression problem described in Eqs.~(\ref{eq:regression_B}) and (\ref{eq:data_bi(t)}), $\Theta_i(t)$ is calculated by linear interpolation as described in Eq.~(\ref{eq:calculation_Theta}), and $\tilde{A}_i(t)$ is obtained from Eq.~(\ref{eq:tilde_A}).
Note that Eq.~(\ref{eq:calculation_varphi}) approximates the slow variable for the spatial phase, $\varphi_i(t)$, as constant over each period of the temporal phase and therefore does not resolve its variation within that period.
Thus, rather than directly calculating the spatial phase as $\hat{c}_i t+\varphi_i(t)$, we use Eq.~(\ref{eq:calculation_Phi}) to obtain $\Phi_i(t)$.

\par 
The temporal phase can be calculated using an alternative method based on the Hilbert transform, without changing the procedure for the spatial phase (see Eqs.~(\ref{eq:A(Phi,Theta)})--(\ref{eq:calculation_Phi})).
Appendix~\ref{appendix:Hilbert} describes the method based on the Hilbert transform and examines how the reconstruction of the phase equations depends on the method used to calculate the temporal phase.

\subsection{Bayesian inference}
\label{subsec:BayesianInference}

\par 
In order to estimate the parameters in Eqs.~(\ref{eq:PhiEquation_expansion}) and (\ref{eq:ThetaEquation_expansion}), we propose a Bayesian inference method.
The Bayesian inference method for noisy limit-torus oscillators is developed by extending the Bayesian framework for reconstructing the phase dynamics of noisy limit-cycle oscillators~\cite{stankovski_inference_2012,duggento_dynamical_2012,luchinsky_inferential_2008,duggento_inferential_2008} to incorporate spatial and temporal phases.
The parameters are estimated from the following dataset:
\begin{align}
    \label{eq:data_D}
    \mathcal{D} = \{ \Phi_{i,n}^\ast, \Theta_{i,n}^\ast, \dot{\Phi}_{i,n}, \dot{\Theta}_{i,n} \}_{i,n}.
\end{align}
The derivatives of the spatial and temporal phases are determined through the midpoint approximation:
\begin{align}
    \label{eq:Phi_ast}
    \Phi_{i,n}^* := \frac{\Phi_i(t_n + \Delta t) + \Phi_i(t_n)}{2}, \\
    \label{eq:Theta_ast}
    \Theta_{i,n}^* := \frac{\Theta_i(t_n + \Delta t) + \Theta_i(t_n)}{2}, \\
    \label{eq:dot_Phi}
    \dot{\Phi}_{i,n} := \frac{\Phi_i(t_n + \Delta t) - \Phi_i(t_n)}{\Delta t}, \\
    \label{eq:dot_Theta}
    \dot{\Theta}_{i,n} := \frac{\Theta_i(t_n + \Delta t) - \Theta_i(t_n)}{\Delta t},
\end{align}
where $n=0,1,\ldots,N-1$ is the data index.
The data need not be temporally successive; the dataset $\mathcal{D}$ can be constructed by assembling data points $(\Phi_{i,n}^*, \Theta_{i,n}^*, \dot{\Phi}_{i,n}, \dot{\Theta}_{i,n})$ drawn from multiple phase time series.
Unlike in the previous subsection, the index $n$ labels data points in $\mathcal{D}$, which need not include all samples taken at every $\Delta t$.

\par 
Here, we introduce vector and matrix notations for the formulation of update rules based on Bayes' theorem.
The time derivatives of the spatial and temporal phases are assembled into a vector as follows:
\begin{align}
    \dot{\boldsymbol{\chi}}_{i,n} 
    := \begin{pmatrix}
        \dot{\Phi}_{i,n} \\
        \dot{\Theta}_{i,n}
    \end{pmatrix}.
\end{align}
The parameter vector is defined as
\begin{align}
    & \boldsymbol{a}_{i}
    :=
    \begin{pmatrix}
        \boldsymbol{a}_i^{\mathrm{s}} \\
        \boldsymbol{a}_i^{\mathrm{t}}
    \end{pmatrix}
    \in \mathbb{C}^{2R},
    & \boldsymbol{a}_i^{\mathrm{p}}
    :=
    \begin{pmatrix}
        a_{i,\boldsymbol{0}}^{\mathrm{p}} \\
        \boldsymbol{a}_{i1}^{\mathrm{p}} \\
        \vdots \\
        \boldsymbol{a}_{i\,i-1}^{\mathrm{p}} \\
        \boldsymbol{a}_{i\,i+1}^{\mathrm{p}} \\
        \vdots \\
        \boldsymbol{a}_{iJ}^{\mathrm{p}}
    \end{pmatrix}
    \in \mathbb{C}^{R},
\end{align}
where 
$
\boldsymbol a_{ij}^{\mathrm p}
:=
\left(
    a_{ij,\boldsymbol m}^{\mathrm p}
\right)_{\boldsymbol m\in\mathbb M}
$. 
The quantity $R := (J-1) \left[(2M_{\mathrm{s}}+1)(2M_{\mathrm{t}}+1) - 1 \right] +1$ is the number of parameters in each phase equation of the spatial and temporal phases.
The vector of Fourier basis functions is given by
\begin{align}
    \boldsymbol{g}_{i,n}
    &:=
    \begin{pmatrix}
        1 \\
        \boldsymbol{g}_{i1,n} \\
        \vdots \\
        \boldsymbol{g}_{i\,i-1,n} \\
        \boldsymbol{g}_{i\,i+1,n} \\
        \vdots \\
        \boldsymbol{g}_{iJ,n}
    \end{pmatrix} \in \mathbb{C}^{R},
\end{align}
where
$
\boldsymbol g_{ij,n}
:= \left( g_{ij,\boldsymbol m,n} \right)_{\boldsymbol m\in\mathbb M}
$ 
with 
\begin{align}
    & g_{ij, \boldsymbol{m}, n}
    := g_{\boldsymbol{m}} (\Phi_{i,n}^* - \Phi_{j,n}^*, \Theta_{i,n}^* - \Theta_{j,n}^*).
\end{align}
We also define the following matrices:
\begin{align}
    & {\rm G}_{i,n}
    :=
    \begin{pmatrix}
    \boldsymbol{g}_{i,n}^\mathrm{T} & 0 \\
    0 & \boldsymbol{g}_{i,n}^\mathrm{T}
    \end{pmatrix} \in \mathbb{C}^{2 \times 2R},
    \\
    &\partial {\rm G}_{i,n}
    :=
    \begin{pmatrix}
    \frac{\partial}{\partial \Phi_i} \boldsymbol{g}_{i,n}^\mathrm{T} & 0 \\
    0 & \frac{\partial}{\partial \Theta_i} \boldsymbol{g}_{i,n}^\mathrm{T}
    \end{pmatrix} \in \mathbb{C}^{2 \times 2R}.
\end{align}
The upper-left and lower-right blocks of the matrix ${\rm G}_{i,n}$ represent the basis functions appearing in the phase equations for the spatial and temporal phases, respectively.
These blocks are identical because both phase equations use the same set of basis functions
(see Eqs.~(\ref{eq:PhiEquation_expansion}) and (\ref{eq:ThetaEquation_expansion})).

\par
Under Gaussian white noise and the midpoint approximation described in Eqs.~(\ref{eq:Phi_ast})--(\ref{eq:dot_Theta}), the likelihood is given by the product over $n$ of the probabilities of observing the step from $(\Phi_i(t_n), \Theta_i(t_n))$ to $(\Phi_i(t_n+\Delta t), \Theta_i(t_n+\Delta t))$.
On the basis of Refs.~\cite{stankovski_inference_2012,duggento_dynamical_2012,luchinsky_inferential_2008,duggento_inferential_2008}, we define the minus log-likelihood function as follows:
\begin{align}
    \label{eq:negative_log_likelihood}
    S_i \left( \mathcal{D} \mid \mathcal{M}_i \right)
    &= \frac{N}{2} \ln |{\rm E}_i|
    + \frac{\Delta t}{2} \sum_{n=0}^{N-1}
    \Biggl[
    \boldsymbol{1}^{\mathrm T}
    \left(\partial {\rm G}_{i,n} \right)
    \boldsymbol{a}_i
    \nonumber\\
    &\qquad
    +
    \left(
    \dot{\boldsymbol{\chi}}_{i,n}
    -
    {\rm G}_{i,n}\, \boldsymbol{a}_i
    \right)^{\dagger}
    {\rm E}_i^{-1}
    \left(
    \dot{\boldsymbol{\chi}}_{i,n}
    -
    {\rm G}_{i,n}\, \boldsymbol{a}_i
    \right)
    \Biggr], 
\end{align}
where $\boldsymbol{1} = (1,1)^\mathrm{T}$. 
Thus, the likelihood function is given by $ L \left( \mathcal{D} \mid \mathcal{M}_i \right) = \exp \left[ - S_i \left( \mathcal{D} \mid \mathcal{M}_i \right) \right]$.
On the basis of Refs.~\cite{stankovski_inference_2012, duggento_dynamical_2012, luchinsky_inferential_2008}, the likelihood includes a correction term arising from the Jacobian of the transformation from the noise to the increments of the phases.
This contribution corresponds to the term $\boldsymbol{1}^{\mathrm T}(\partial {\rm G}_{i,n})\boldsymbol a_i$ in Eq.~(\ref{eq:negative_log_likelihood}), where the Jacobian is approximated by the product of its diagonal terms~\cite{stankovski_inference_2012, duggento_dynamical_2012}.
By the conjugate symmetry of the Fourier coefficients, Eq.~(\ref{eq:negative_log_likelihood}) is real-valued.

\par
To reconstruct the phase equations, the unknown model parameters in Eqs.~(\ref{eq:PhiEquation_expansion}) and (\ref{eq:ThetaEquation_expansion}), denoted by 
\begin{align}
    \label{eq:parameter_M}
    \mathcal{M}_i 
    =
    \{
        \boldsymbol{a}_i,
        {\rm E}_i
    \},
\end{align}
are estimated from the data $\mathcal{D}$.
We assume a prior distribution for the parameter vector $\boldsymbol{a}_i$, whereas the noise covariance matrix ${\rm E}_i$ is determined solely by the likelihood, without specifying a prior distribution.
The prior distribution for $\boldsymbol{a}_i$ is given by a Gaussian distribution~\cite{note:gaussian_prior}, i.e.,
\begin{align}
    \label{eq:prior_distribution}
    \boldsymbol{a}_{i} \sim \mathcal{N}(\boldsymbol{a}_{i,\mathrm{prior}}, \Sigma_{i,\mathrm{prior}}).
\end{align}
Given the prior distribution, which encodes knowledge about the unknown parameters, together with the likelihood function, we determine the posterior distribution in accordance with Bayes' theorem:
\begin{align}
    \label{eq:bayes}
    P_{\mathrm{post}} \left( \mathcal{M}_i \mid \mathcal{D} \right)
    \propto 
    L \left( \mathcal{D} \mid \mathcal{M}_i \right)
    P_{\mathrm{prior}} \left( \mathcal{M}_i \right),
\end{align}
where $P_{\mathrm{prior}} \left( \mathcal{M}_i \right)$ and $P_{\mathrm{post}} \left( \mathcal{M}_i \mid \mathcal{D} \right)$ denote the prior and posterior distributions of the parameters, respectively.
For fixed ${\rm E}_i$, the Gaussian prior distribution for $\boldsymbol{a}_{i}$ is conjugate to the likelihood; hence, the conditional posterior distribution of $\boldsymbol{a}_{i}$ is also a Gaussian distribution.
Eq.~(\ref{eq:bayes}) is written in terms of $\mathcal{M}_i$ for notational convenience; no prior distribution is specified for ${\rm E}_i$.

\par
The parameter vector $\boldsymbol{a}_i$ is obtained by maximizing the posterior, whereas ${\rm E}_i$ is obtained by minimizing the negative log-likelihood~\cite{stankovski_inference_2012,duggento_dynamical_2012,luchinsky_inferential_2008,duggento_inferential_2008}.
The model parameters are determined by the following equations:
\begin{align}
    \label{eq:update_E}
    & {\rm E}_i
    = \frac{\Delta t}{N} \sum_{n=0}^{N-1}
    \bigl( 
        \dot{\boldsymbol{\chi}}_{i,n} - {\rm G}_{i,n}\, \boldsymbol{a}_i
    \bigr)
    \bigl( 
        \dot{\boldsymbol{\chi}}_{i,n} - {\rm G}_{i,n}\, \boldsymbol{a}_i
    \bigr)^\dagger, \\
    \label{eq:update_Xi}
    & \Xi_i
    = \Xi_{i, \mathrm{prior}}
    + \Delta t \sum_{n=0}^{N-1} {\rm G}_{i,n}^\dagger {\rm E}_i^{-1} {\rm G}_{i,n}, \\
    \label{eq:update_r}
    & \boldsymbol{r}_i
    = \Xi_{i, \mathrm{prior}}\, \boldsymbol{a}_{i, \mathrm{prior}}
    + \Delta t \sum_{n=0}^{N-1} {\rm G}_{i,n}^\dagger {\rm E}_i^{-1} \dot{\boldsymbol{\chi}}_{i,n} \nonumber \\
    & \quad \quad - \frac{\Delta t}{2} \sum_{n=0}^{N-1} (\partial {\rm G}_{i,n})^{\dagger} \boldsymbol{1}, \\
    \label{eq:update_a}
    & \boldsymbol{a}_i
    = \Xi_i^{-1}
    \boldsymbol{r}_i,
\end{align}
where $\Xi_i = \Sigma_i^{-1}$ and $\Xi_{i, \mathrm{prior}} = \Sigma_{i, \mathrm{prior}}^{-1}$.
Starting from $\boldsymbol{a}_i=\boldsymbol{a}_{i,\mathrm{prior}}$ and $\Sigma_i = \Sigma_{i,\mathrm{prior}}$, we recursively calculate Eqs.~(\ref{eq:update_E})--(\ref{eq:update_a}) until the stationary point of $S_i \left( \mathcal{D} \mid \mathcal{M}_i \right)$ is obtained.

\par 
The initial conditions
$\boldsymbol{a}_{i,\mathrm{prior}}$ and
$\Sigma_{i,\mathrm{prior}}$
are chosen to satisfy the conjugate-symmetry conditions
\begin{align}
    & a_{ij,-\boldsymbol{m}}^{\mathrm p}
    =
    \left(
        a_{ij,\boldsymbol{m}}^{\mathrm p}
    \right)^\dagger, \\
    & \left[
        \Sigma_{i}^{\mathrm p\mathrm q}
    \right]_{(j,-\boldsymbol{m}),(k,-\boldsymbol{m}')}
    =
    \left(
        \left[
            \Sigma_{i}^{\mathrm p\mathrm q}
        \right]_{(j,\boldsymbol{m}),(k,\boldsymbol{m}')}
    \right)^\dagger, \\
    & \left[\Sigma_i^{\mathrm p\mathrm q}\right]_{\boldsymbol{0},(k,-\boldsymbol m')}
    =
    \left(\left[\Sigma_i^{\mathrm p\mathrm q}\right]_{\boldsymbol{0},(k,\boldsymbol m')}\right)^\dagger, \\
    & \left[\Sigma_i^{\mathrm p\mathrm q}\right]_{(j,-\boldsymbol m),\boldsymbol{0}}
    =
    \left(\left[\Sigma_i^{\mathrm p\mathrm q}\right]_{(j,\boldsymbol m),\boldsymbol{0}}\right)^\dagger,
\end{align}
where $\mathrm p,\mathrm q\in\{\mathrm s,\mathrm t\}$ and $\boldsymbol{m},\boldsymbol{m}'\in\mathbb{M}$.
Here, $\cdot^\dagger$ denotes complex conjugation when acting on a scalar.
The entry $\left[\Sigma_i^{\mathrm p\mathrm q}\right]_{(j,\boldsymbol m),(k,\boldsymbol m')}$ denotes the covariance between $a_{ij,\boldsymbol m}^{\mathrm p}$ and $a_{ik,\boldsymbol m'}^{\mathrm q}$. 
For convenience, we also use this notation for covariances involving the constant coefficients.
When $(j,\boldsymbol m)$ is replaced by $\boldsymbol0$, we use $a_{i,\boldsymbol0}^{\mathrm p}$ instead of $a_{ij,\boldsymbol m}^{\mathrm p}$; similarly, when $(k,\boldsymbol m')$ is replaced by $\boldsymbol0$, we use $a_{i,\boldsymbol0}^{\mathrm q}$ instead of $a_{ik,\boldsymbol m'}^{\mathrm q}$.
Furthermore, the initial conditions satisfy $a_{i,\boldsymbol0}^{\mathrm p}\in\mathbb R$ and $\left[\Sigma_i^{\mathrm p\mathrm q}\right]_{\boldsymbol0,\boldsymbol0}\in\mathbb R$, since $a_{i,\boldsymbol0}^{\mathrm s}$ and $a_{i,\boldsymbol0}^{\mathrm t}$ correspond to the effective traveling velocity and oscillation frequency, respectively.
Given the relation $g_{-\boldsymbol{m}}=g_{\boldsymbol{m}}^\dagger$ and the fact that the vectors $\dot{\boldsymbol{\chi}}_{i,n}$ are real-valued, the updates in Eqs.~(\ref{eq:update_E})--(\ref{eq:update_a}) preserve these conjugate-symmetry conditions for $\boldsymbol{a}_i$ and $\Sigma_i=\Xi_i^{-1}$.


\par
The Bayesian inference formulation includes nonzero off-diagonal entries of ${\rm E}_i$. 
However, when the primary purpose is to reconstruct the deterministic part of the phase equations, the covariance between the noise terms in the spatial and temporal phases may be neglected, provided that doing so does not degrade the reconstruction accuracy.
Under this condition, one can impose $E_i^{\mathrm{st}}=E_i^{\mathrm{ts}}=0$, thereby simplifying the inference problem.
In Appendix~\ref{appendix:IndependentInference}, we examine whether the phase equations can be accurately reconstructed under this simplification.
The results remain unchanged, except that $E^{\mathrm{st}}_i = E^{\mathrm{ts}}_i = 0$.


\section{Application to simulation data of traveling breathers in coupled Gray-Scott models}
\label{sec:sec3}

\par
In this section, we demonstrate the reconstruction of the phase equations from simulation data of traveling breathers in coupled Gray-Scott models.
In Sec.~\ref{subsec:Gray-Scott}, we describe the coupled Gray-Scott models used to generate the simulation data.
The phase equations obtained through the model-driven approach in Ref.~\cite{arai_phase_2025} are used as the ground truth.
In Sec.~\ref{subsec:PhaseTimeSeries_of_TB}, we illustrate the calculation of the spatial and temporal phases.
In Sec.~\ref{subsec:Results_of_BayesianInference}, we present the results of the Bayesian inference.

\subsection{Traveling breathers in coupled Gray-Scott models}
\label{subsec:Gray-Scott}

\par
We consider a pair of coupled Gray-Scott models ($J=2$) described by
\begin{align}
    \label{eq:GrayScottModel_X}
    & \boldsymbol{X}_i(x, t)
    = \begin{pmatrix}
    u_i \\
    v_i
    \end{pmatrix}, \\
    \label{eq:GrayScottModel_F}
    & \boldsymbol{F}_i(\boldsymbol{X}_i(x,t))
    =
    \begin{pmatrix}
    u_i^2 v_i - (f + k) u_i \\
    - u_i^2 v_i + f (1 - v_i) 
    \end{pmatrix}, \\
    \label{eq:GrayScottModel_G}
    & \epsilon \boldsymbol{G}_{ij}(\boldsymbol{X}_i(x,t), \boldsymbol{X}_j(x,t))
    = \epsilon (\boldsymbol{X}_j(x,t) - \boldsymbol{X}_i(x,t)),
\end{align}
where $u_i:=u_i(x,t)$ and $v_i:=v_i(x,t)$.
The system has a size of $2L = 250$, which is sufficiently large compared to that of the breather. 
The diffusion coefficient is ${\rm D}_i = \mathrm{diag}(1.0, 1.9)$ for $i=1,2$ and the coupling intensity is $\epsilon = 1.0 \times 10^{-6}$. 
The other parameters are set to $f = 0.018$ and $k = 0.052$~\cite{arai_phase_2025,yadome_chaotic_2011}.

\par
\begin{figure*}[!t]
    \begin{minipage}[t]{0.48\textwidth}
        \begin{center}
            \includegraphics[scale=0.95]{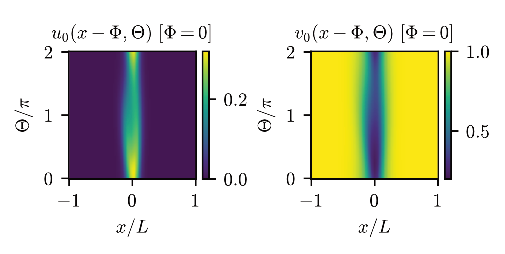}
            \caption{
                Limit-torus solution,
                $\boldsymbol{X}_0(x-\Phi,\Theta)
                = \left(
                    u_0(x-\Phi,\Theta), v_0(x-\Phi,\Theta)
                    \right)$
                with $\Phi = 0$,
                for the Gray-Scott model.
            }
            \label{fig:fig2}
        \end{center}
    \end{minipage}
    \hfill
    \begin{minipage}[t]{0.48\textwidth}
        \begin{center}
            \includegraphics[scale=0.9]{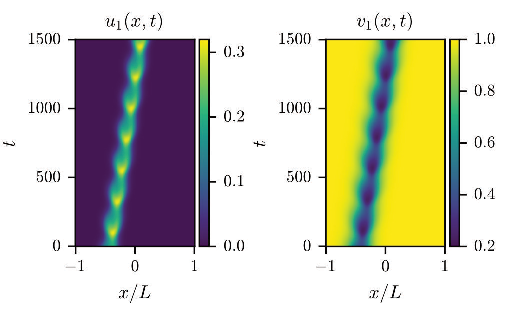}
            \caption{
                Example of time series
                $\boldsymbol{X}_1(x, t) = \left( u_1(x, t), v_1(x, t) \right)$
                of the Gray-Scott model ($\epsilon = 0$).
            }
            \label{fig:fig3}
        \end{center}
    \end{minipage}
\end{figure*}

\begin{figure}[!t]
    \begin{center}
        \includegraphics[scale=0.95]{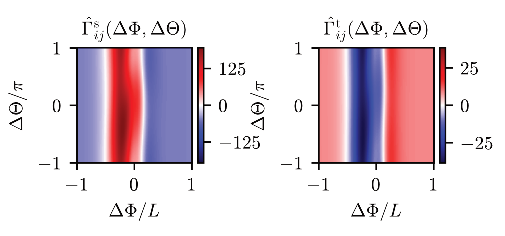}
        \caption{
            Phase coupling functions,
            $\hat{\Gamma}_{ij}^{\mathrm{s}}(\Delta \Phi, \Delta \Theta)$ and $\hat{\Gamma}_{ij}^{\mathrm{t}}(\Delta \Phi, \Delta \Theta)$, for the coupled Gray-Scott models.
        }
        \label{fig:fig4}
    \end{center}
\end{figure}

 
Figure~\ref{fig:fig2} shows the limit-torus solution $\boldsymbol{X}_0(x-\Phi_i, \Theta_i)$ with $\Phi_i = 0$. 
The spatiotemporal pattern of $\boldsymbol{X}_0(x-\Phi_i, \Theta_i)$ breaks the spatial reflection symmetry.
We determine the state corresponding to $\Phi_i = 0$ in the limit-torus solution by requiring that it satisfy the condition in Eq.~(\ref{eq:B_condition}), where
\begin{align}
    B(\Theta_i) = \frac{L}{\pi}\arg A(0, \Theta_i).
\end{align}
This expression for $B(\Theta)$ follows from substituting $\Phi_i = 0$ into Eq.~(\ref{eq:arg_A(Phi,Theta)}).
Figure~\ref{fig:fig3} shows the time series $\boldsymbol{X}_1(x, t)$ in the absence of coupling and noise ($\epsilon = 0$, $\boldsymbol{\xi}_i(x,t) = \boldsymbol{0}$).
The traveling breather exhibits spatial and temporal phases that increase linearly in time, i.e., $\Phi_i = c_i t$ and $\Theta_i = \omega_i t$, respectively.
In simulations of the coupled Gray-Scott models, the velocities of the spatial and temporal phases are modulated as described by the phase coupling functions.
Figure~\ref{fig:fig4} shows the phase coupling functions $\hat{\Gamma}_{ij}^{\mathrm{s}}(\Delta \Phi, \Delta \Theta)$ and $\hat{\Gamma}_{ij}^{\mathrm{t}}(\Delta \Phi, \Delta \Theta)$.
As the two Gray-Scott models ($i=1,2$) are identical, they have the same limit-torus solution.
Furthermore, because the coupling terms between the identical systems are symmetric, the phase coupling functions $\hat{\Gamma}_{ij}^{\mathrm{s}}(\Delta \Phi, \Delta \Theta)$ and $\hat{\Gamma}_{ij}^{\mathrm{t}}(\Delta \Phi, \Delta \Theta)$ are identical for $(i,j)=(1,2)$ and $(2,1)$.

\par
\begin{figure*}[!t]
    \begin{center}
        \includegraphics[scale=0.95]{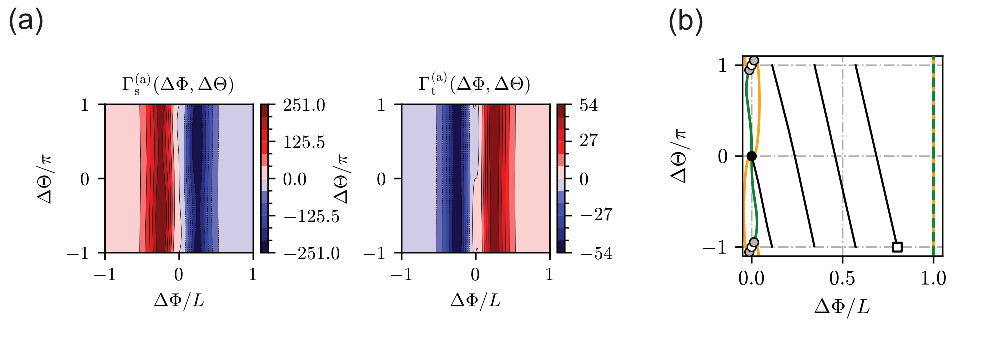}
        \caption{
            Dynamics of spatial and temporal phase differences for the coupled Gray-Scott models in the absence of noise.
            (a)~
            Antisymmetric components of the phase coupling functions,
            $\Gamma_{\mathrm{s}}^{(\mathrm{a})}(\Delta \Phi, \Delta \Theta)$ and
            $\Gamma_{\mathrm{t}}^{(\mathrm{a})}(\Delta \Phi, \Delta \Theta)$,
            which determine the time evolution of the phase differences.
            (b)~
            Nullclines of
            $\Gamma_{\mathrm{s}}^{(\mathrm{a})}(\Delta \Phi, \Delta \Theta)$ (green line)
            and
            $\Gamma_{\mathrm{t}}^{(\mathrm{a})}(\Delta \Phi, \Delta \Theta)$ (orange line).
            Closed and open circles indicate the stable and unstable fixed points,
            respectively, and gray circles indicate the saddle points.
            The black line traces the trajectory starting from $(\Delta\Phi/L,\Delta\Theta/\pi)=(0.8, -1)$ (open square) and converging to $(0,0)$.
            It is obtained by numerically integrating Eqs.~(\ref{eq:d(Delta_Phi)/dt}) and (\ref{eq:d(Delta_Theta)/dt}).
        }
        \label{fig:fig5}
    \end{center}
\end{figure*}

We define the spatial and temporal phase differences as $\Delta \Phi := \Phi_1 - \Phi_2$ and $\Delta \Theta := \Theta_1 - \Theta_2$, respectively.
In the absence of noise ($\boldsymbol{\xi}_i(x,t)=\boldsymbol{0}$ and hence $\boldsymbol{\eta}_i(t)=\boldsymbol{0}$), the time evolution of the phase differences is given as follows:
\begin{align}
    \label{eq:d(Delta_Phi)/dt}
    \frac{\mathrm{d}}{\mathrm{d}t} \, \Delta \Phi(t)
    & = \epsilon \, \Gamma_\mathrm{s}^{(\mathrm{a})}(\Delta \Phi, \Delta \Theta),  \\
    \label{eq:d(Delta_Theta)/dt}
    \frac{\mathrm{d}}{\mathrm{d}t} \, \Delta \Theta(t)
    & = \epsilon \, \Gamma_\mathrm{t}^{(\mathrm{a})}(\Delta \Phi, \Delta \Theta),
\end{align}
where
\begin{align}
    \label{eq:Gamma_s^a}
    \Gamma_\mathrm{s}^{(\mathrm{a})}(\Delta \Phi, \Delta \Theta)
    &= \Gamma_{12}^{\mathrm{s}}(\Delta \Phi, \Delta \Theta)
    - \Gamma_{21}^{\mathrm{s}}(-\Delta \Phi, -\Delta \Theta), \\
    \label{eq:Gamma_t^a}
    \Gamma_\mathrm{t}^{(\mathrm{a})}(\Delta \Phi, \Delta \Theta)
    &= \Gamma_{12}^{\mathrm{t}}(\Delta \Phi, \Delta \Theta)
    - \Gamma_{21}^{\mathrm{t}}(-\Delta \Phi, -\Delta \Theta).
\end{align}
These equations are obtained from Eqs.~(\ref{eq:PhaseEquation_Phi}) and (\ref{eq:PhaseEquation_Theta}) with $c_1 = c_2$ and $\omega_1 = \omega_2$.
The functions $\Gamma_\mathrm{s}^{(\mathrm{a})}$ and $\Gamma_\mathrm{t}^{(\mathrm{a})}$ are the antisymmetric components of the phase coupling functions because the original phase coupling functions are identical.
Figure~\ref{fig:fig5}(a) shows $\Gamma_\mathrm{s}^{(\mathrm{a})}(\Delta \Phi, \Delta \Theta)$ and $\Gamma_\mathrm{t}^{(\mathrm{a})}(\Delta \Phi, \Delta \Theta)$. 
Using these functions, we can identify the fixed points and analyze their linear stability in the $(\Delta \Phi, \Delta \Theta)$ space.
Figure~\ref{fig:fig5}(b) shows the nullclines of $\Gamma_\mathrm{s}^{(\mathrm{a})}(\Delta \Phi, \Delta \Theta)$ and $\Gamma_\mathrm{t}^{(\mathrm{a})}(\Delta \Phi, \Delta \Theta)$, along with the locations of the fixed points and a typical trajectory.
Under the phase reduction approximation, only $(\Delta\Phi,\Delta\Theta)=(0,0)$, which represents the spatial and temporal in-phase state, is linearly stable~\cite{note:in_phase_stability}.


\par
Time series of $\boldsymbol{X}_i(x,t)$ were generated through numerical simulations of the coupled Gray-Scott models.
Time integration was performed using the explicit Euler scheme with a time step of $0.01$, and the system was discretized using a spatial grid size of $2L/2^{10}$.
We sampled $u_i(x,t)$ with a sampling interval of $\Delta t = 1.0$.
The initial conditions were given by $\boldsymbol{X}_i(x,0)=\boldsymbol{X}_0(x-\Phi_i,\Theta_i)$ with specified values of $\Delta \Phi / L$ and $\Delta \Theta / \pi$.
The numerical simulations were performed for the following noise intensities:
$\sigma_i^2 =
2.5\times10^{-13},\,
1.0\times10^{-12},\,
4.0\times10^{-12},\,
1.6\times10^{-11},\,
1.0\times10^{-10},\,
4.0\times10^{-10},\,
1.6\times10^{-9},\,$
and $2.5\times10^{-9}$.

\par 
In the numerical simulations, to cover the $(\Delta \Phi / L, \Delta \Theta / \pi)$ plane, we generated multiple long and short time series of $\boldsymbol{X}_i(x,t)$ as follows.
For the long time series, a total of 64 trajectories were generated from initial conditions with $\Delta \Phi / L = \pm 0.8$ and $\Delta \Theta / \pi = \frac{1}{16} \ell$ $(\ell = 0,1,\ldots,31)$. 
These trajectories were computed up to $\epsilon t = t_{\mathrm{max}}$, where $t_{\mathrm{max}}$ ranges from $7.0$ to $10.0$.
For sufficiently weak noise, all 64 trajectories converge to the in-phase synchronized state, i.e., $(\Delta\Phi,\Delta\Theta)=(0,0)$, consistent with the nullclines shown in Fig.~\ref{fig:fig5}(b).
Furthermore, to cover the regions of $(\Delta \Phi / L, \Delta \Theta / \pi)$ with $|\Delta \Phi / L| > 0.8$, which are not covered by the long time series, we generated a large number of short time series.
In this region, the phase differences evolve slowly and remain close to their initial values even over long simulations. 
Therefore, instead of performing long simulations, we densely sampled the initial conditions to generate a large number of short time series.
For the short time series, the initial conditions were given by $\Delta \Phi / L \simeq \pm 0.82, \pm 0.84, \ldots, \pm 0.98$ and $\Delta \Theta / \pi = \frac{1}{16} \ell$ $(\ell=0,1,\ldots,31)$, and the corresponding trajectories were computed up to $\epsilon t = 0.1$.

\subsection{Phase time series of traveling breathers}
\label{subsec:PhaseTimeSeries_of_TB}

\par
\begin{figure}[!tbp]
    \begin{center}
        \includegraphics[scale=0.95]{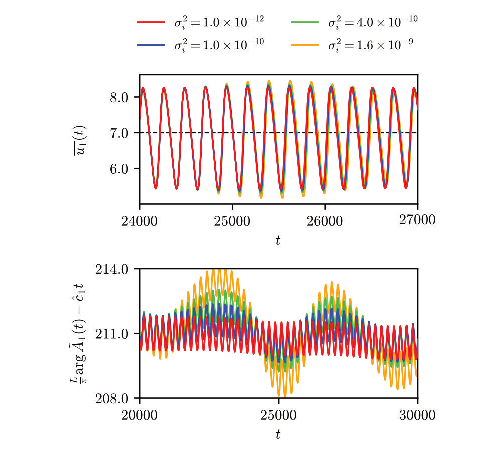}
        \caption{
            Time series of $\overline{u_1}(t)$ (top) and
            $\frac{L}{\pi}\arg \tilde{A}_1(t)-\hat{c}_1 t$ (bottom).
            The red, blue, green, and orange curves correspond to
            $\sigma_i^2 = 1.0 \times 10^{-12}$,
            $1.0 \times 10^{-10}$,
            $4.0 \times 10^{-10}$, and
            $1.6 \times 10^{-9}$, respectively.
            In the top panel, the dashed line indicates the Poincar\'e section at $\overline{u_1}(t)=7.0$.
            In the bottom panel, the value $\hat{c}_1 = 4.30057 \times 10^{-2}$, obtained for the Gray-Scott model without noise and coupling ($\sigma_i^2=0$, $\epsilon=0$), is used as a reference.
        }
        \label{fig:fig6}
    \end{center}
\end{figure}

We calculated the corresponding spatial and temporal phase time series from $u_i(x,t)$ obtained from numerical simulations.
Figure~\ref{fig:fig6} displays the time series of $\overline{u_1}(t)$ and $\frac{L}{\pi} \arg \tilde{A}_1(t) - \hat{c}_1 t$ for different noise intensities, all starting from the same initial condition. 
The temporal phase time series was obtained by linear interpolation as described in Eq.~(\ref{eq:calculation_Theta}). 
In this procedure, we set the Poincar\'e section at $\overline{u_i}(t) = 7.0$ and defined $t_{i,k}$ as the time at which $\overline{u_i}(t)$ crosses this section with a positive slope.
According to Eq.~(\ref{eq:increment_argA}), the time series $\frac{L}{\pi} \arg \tilde{A}_i(t) - \hat{c}_i t$ consists of a periodic function $\tilde{B}_i(\Theta_i(t))$ and a slow variable $\varphi_i(t)$.

\par
\begin{figure}[!tbp]
    \begin{center}
        \includegraphics[scale=0.95]{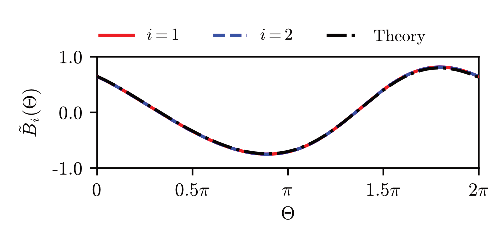}
        \caption{
            Periodic functions $\tilde{B}_i(\Theta)$ obtained from the regression described in Eqs.~(\ref{eq:PeriodicFunction_B})--(\ref{eq:data_bi(t)}).
            The solid red and dashed blue curves correspond to $i=1$ and $i=2$,
            respectively.
            We show $\tilde{B}_i(\Theta)$ obtained for $\sigma_i^2 = 1.0 \times 10^{-12}$ as a representative example; similar results are obtained for other noise intensities.
            The dash-dotted black curve indicates the theoretical function
            $B(\Theta)$.
        }
        \label{fig:fig7}
    \end{center}
\end{figure}

On the basis of Eqs.~(\ref{eq:Regression_hatc}) and (\ref{eq:PeriodicFunction_B})--(\ref{eq:data_bi(t)}), we estimated $\hat{c}_i$ and the function $\tilde{B}_i(\Theta_i)$ from all time series of $\Theta_i(t)$ and $\tilde{A}_i(t)$.
Figure~\ref{fig:fig7} shows that $\tilde{B}_i(\Theta_i)$ obtained for $\sigma_i^2 = 1.0 \times 10^{-12}$ coincides with $B(\Theta_i)$ obtained from the limit-torus solution; moreover, $\tilde{B}_i(\Theta_i)$ is essentially identical for all values of $\sigma_i^2$.
The time series of $\tilde{B}_i(\Theta_i(t))$ accounts for the fast oscillation in $\frac{L}{\pi} \arg \tilde{A}_i(t) - \hat{c}_i t$ (Fig.~\ref{fig:fig6}).

\par
\begin{figure}[!tbp]
    \begin{center}
        \includegraphics[scale=0.95]{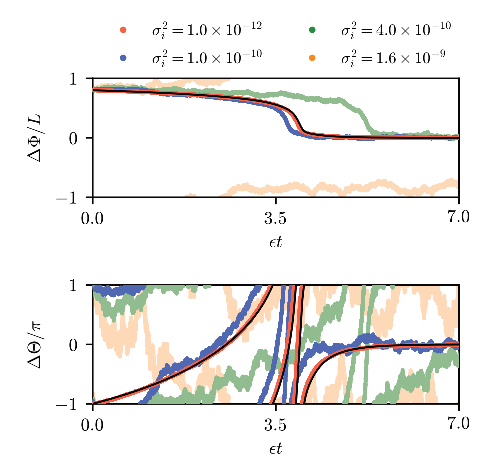}
        \caption{
            Time series of the phase differences $\Delta \Phi(t)$ (top) and $\Delta \Theta(t)$ (bottom) as functions of $\epsilon t$.
            The colored curves show the phase differences calculated from Eqs.~(\ref{eq:calculation_Theta}) and (\ref{eq:calculation_Phi}).
            The red, blue, green, and orange curves correspond to $\sigma_i^2 = 1.0 \times 10^{-12}$,
            $1.0 \times 10^{-10}$,
            $4.0 \times 10^{-10}$, and
            $1.6 \times 10^{-9}$, respectively.
            The black curves indicate the time evolution of the phase differences obtained by numerically integrating Eqs.~(\ref{eq:d(Delta_Phi)/dt}) and (\ref{eq:d(Delta_Theta)/dt}).
            The initial phase differences are $(\Delta\Phi/L,\Delta\Theta/\pi)=(0.8,-1)$.
        }
        \label{fig:fig8}
    \end{center}
\end{figure}

 
Once $\tilde{B}_i(\Theta_i(t))$ is obtained, the spatial phase is calculated as described in Eq.~(\ref{eq:calculation_Phi}).
Figure~\ref{fig:fig8} shows the time series of the phase differences. 
For $\sigma_i^2 \leq 1.0 \times 10^{-10}$, the spatiotemporal phase dynamics converge to the in-phase synchronized state and follow the theoretical time evolution obtained from Eqs.~(\ref{eq:d(Delta_Phi)/dt})--(\ref{eq:Gamma_t^a}); otherwise, they do not converge and show large deviations from the theoretical time evolution.

\subsection{Results of the Bayesian inference}
\label{subsec:Results_of_BayesianInference}
 
\par
We performed Bayesian inference to estimate the model parameters $\mathcal{M}_i$. 
The dataset $\mathcal{D}$ used for the estimation was constructed as follows.
Each data point $(\Phi_{i,n}^{\ast}, \Theta_{i,n}^{\ast}, \dot{\Phi}_{i,n}, \dot{\Theta}_{i,n})$ was computed based on Eqs.~(\ref{eq:Phi_ast})--(\ref{eq:dot_Theta}). 
The $(\Delta \Phi / L, \Delta \Theta / \pi)$ space was divided into sections of size $0.05 \times 0.05$, and 250 data points were sampled from each section such that the phase differences $(\Phi_{1,n}^{\ast} - \Phi_{2,n}^{\ast},\, \Theta_{1,n}^{\ast} - \Theta_{2,n}^{\ast})$ fell within the corresponding section.
The phase coupling functions were expanded up to $M_\mathrm{s}=20$ and $M_\mathrm{t}=5$ harmonics; thus, the dimension of the parameter vector $\boldsymbol{a}_i$ was $2R=902$.
The hyperparameters of the prior distribution were set to $\boldsymbol{a}_{i,\mathrm{prior}}=\boldsymbol{0}$ and $\Sigma_{i,\mathrm{prior}}=1.0 \times 10^6 {\rm I}$.

\begin{figure*}[!t]
    \begin{center}
        \includegraphics[scale=0.95]{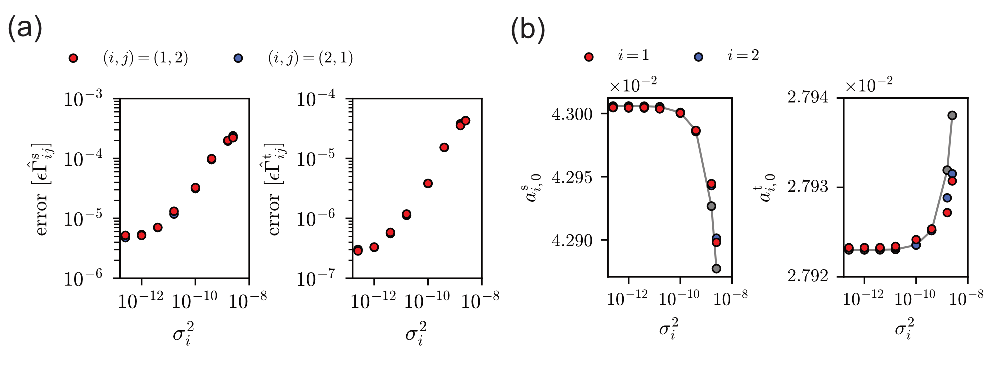}
        \caption{
            Results of the Bayesian inference for the phase coupling functions and the constant terms versus the noise intensity $\sigma_i^2$.
            (a)~ 
            The prediction errors of the phase coupling functions $\epsilon \hat{\Gamma}_{ij}^{\mathrm{s}}$ and $\epsilon \hat{\Gamma}_{ij}^{\mathrm{t}}$.
            The red and blue circles correspond to $(i,j)=(1,2)$ and $(2,1)$, respectively.
            The error is defined by Eqs.~(\ref{eq:prediction_error}) and (\ref{eq:prediction_error2}).
            Both axes are shown on logarithmic scales (log-log plot).
            (b)~
            The constant terms, $a_{i, \mathbf{0}}^{\mathrm{s}}$ and $a_{i, \mathbf{0}}^{\mathrm{t}}$. 
            The red and blue circles correspond to $i=1$ and $i=2$, respectively.
            The gray circles indicate the reference values constructed from numerical simulations of the uncoupled Gray-Scott model~\cite{note:true_constant_terms}.
            The constant terms vary with $\sigma_i^2$ owing to noise-induced shifts.
            The horizontal axis is shown on a logarithmic scale.
        }
        \label{fig:fig9}
    \end{center}
\end{figure*}

\begin{figure*}[!t]
    \begin{center}
        \includegraphics[scale=0.95]{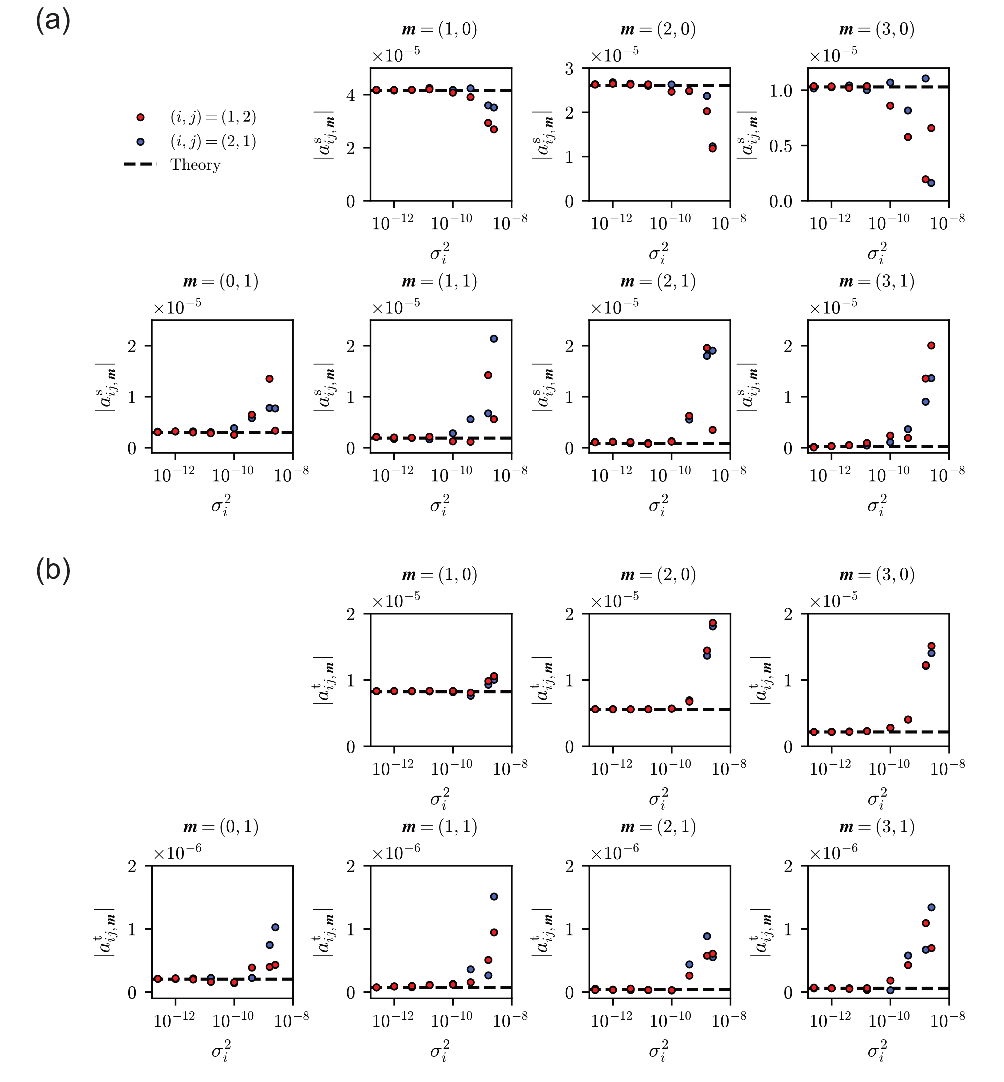}
        \caption{
            Results of the Bayesian inference for the Fourier coefficients of the phase coupling functions versus the noise intensity $\sigma_i^2$.
            The red and blue circles correspond to $(i,j)=(1,2)$ and $(2,1)$, respectively.
            The horizontal axis is shown on a logarithmic scale.
            (a)~
            The Fourier coefficients, $|a_{ij,\boldsymbol{m}}^{\mathrm{s}}|$ with $\boldsymbol{m} = (1,0), (2,0), (3,0), (0,1), (1,1), (2,1)$, and $(3,1)$.
            The gray dashed horizontal lines indicate the theoretical values, obtained from the Fourier expansion of $\hat{\Gamma}_{ij}^{\mathrm{s}}(\Delta \Phi, \Delta \Theta)$.
            (b)~
            The same as panel (a) but for the Fourier coefficients $|a_{ij,\boldsymbol{m}}^{\mathrm{t}}|$.
            The gray dashed horizontal lines indicate the theoretical values, obtained from the Fourier expansion of $\hat{\Gamma}_{ij}^{\mathrm{t}}(\Delta \Phi, \Delta \Theta)$.
        }
        \label{fig:fig10}
    \end{center}
\end{figure*}

\begin{figure}[!tbp]
    \begin{center}
        \includegraphics[scale=0.95]{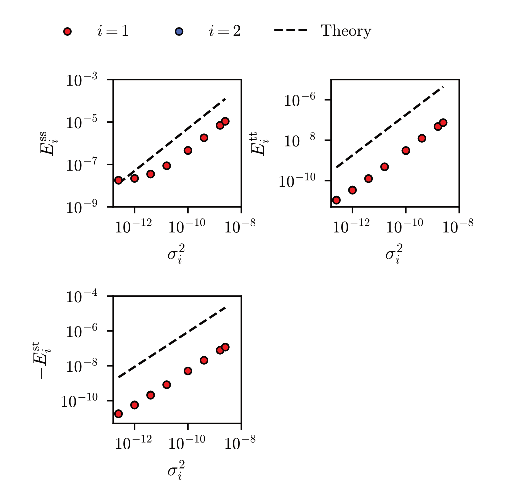}
        \caption{
            Results of the Bayesian inference for the covariance matrix ${\rm E}_i$ versus the noise intensity $\sigma_i^2$.
            The top-left, top-right, and bottom panels show
            $E_i^{\mathrm{s}\mathrm{s}}$, $E_i^{\mathrm{t}\mathrm{t}}$, and
            $-E_i^{\mathrm{s}\mathrm{t}}$, respectively.
            The red and blue circles correspond to $i=1$ and $i=2$, respectively.
            The dashed lines indicate the theoretical values, which follow
            $|E_{i}^{\mathrm{pq}}| \propto \sigma_i^2$ for $\mathrm{p,q} \in \{\mathrm{s},\mathrm{t}\}$
            (see Appendix~\ref{appendix:noise_derivation}).
            Both axes are shown on logarithmic scales (log-log plot).
        }
        \label{fig:fig11}
    \end{center}
\end{figure}

\par
Figure~\ref{fig:fig9}(a) shows the prediction errors of the phase coupling functions, $\epsilon \hat{\Gamma}_{ij}^{\mathrm{s}}$ and $\epsilon \hat{\Gamma}_{ij}^{\mathrm{t}}$, as functions of $\sigma_i^2$. 
The prediction error is defined as 
\begin{align}
    \label{eq:prediction_error}
    & \mathrm{error} \left[ \epsilon \hat{\Gamma}_{ij}^{\mathrm{p}}  \right]
    \nonumber \\
    & = \left(
    \frac{1}{4\pi L}
    \int_{-L}^{L} \, \mathrm{d}\Delta \Phi
    \int_{-\pi}^{\pi} \, \mathrm{d}\Delta \Theta \,
    \left[
        W_{ij}^{\mathrm{p}}(\Delta \Phi,\Delta \Theta)
    \right]^2
    \right)^{1/2},
    \\  
    \label{eq:prediction_error2}
    & W_{ij}^{\mathrm{p}}(\Delta \Phi,\Delta \Theta)
        := X_{ij}^{\mathrm{p}}(\Delta \Phi,\Delta \Theta)
        - Y_{ij}^{\mathrm{p}}(\Delta \Phi,\Delta \Theta),
\end{align}
where $X_{ij}^{\mathrm{p}}(\Delta \Phi,\Delta \Theta)$ and $Y_{ij}^{\mathrm{p}}(\Delta \Phi,\Delta \Theta)$ represent the estimated and true $\epsilon \hat{\Gamma}_{ij}^{\mathrm{p}}(\Delta \Phi,\Delta \Theta)$, respectively.
In our numerical simulations, we set $\sigma_i^2$ to be the same for all $i$; thus, the prediction error defined by Eqs.~(\ref{eq:prediction_error}) and (\ref{eq:prediction_error2}) is treated as a function of $\sigma_i^2$.
For $\sigma_i^2 \gtrsim 4.0 \times 10^{-10}$, the prediction errors for the spatial and temporal phase coupling functions are of order $\mathcal{O}(10^{-4})$ and $\mathcal{O}(10^{-5})$, respectively, comparable to the magnitudes of $\epsilon \hat{\Gamma}_{ij}^{\mathrm{s}}$ and $\epsilon \hat{\Gamma}_{ij}^{\mathrm{t}}$, indicating that the reconstruction is no longer reliable.
For $\sigma_i^2 \leq 1.0 \times 10^{-10}$, the prediction errors are significantly smaller than these values, indicating that the phase coupling functions are accurately reconstructed.
Figure~\ref{fig:fig9}(b) shows the estimated constant terms together with the corresponding true values~\cite{note:true_constant_terms}.
The deviations of the estimates of $a_{i,\boldsymbol{0}}^{\mathrm{s}}$ and $a_{i,\boldsymbol{0}}^{\mathrm{t}}$ from the true values increase for $\sigma_i^2 \gtrsim 4.0 \times 10^{-10}$.
Nevertheless, the noise-induced shifts are qualitatively captured.
For $\sigma_i^2 \leq 1.0 \times 10^{-10}$, the estimates agree well with the true values.
As shown in Figs.~\ref{fig:fig9}(a) and~\ref{fig:fig9}(b), the deterministic part of the phase equations is accurately reconstructed in the weak-noise regime ($\sigma_i^2 \leq 1.0 \times 10^{-10}$).

\par 
To examine these results in more detail, we compared the estimated values of individual Fourier coefficients with the theoretical values.
Figures~\ref{fig:fig10}(a) and~\ref{fig:fig10}(b) show the estimated Fourier coefficients $|a_{ij, \boldsymbol{m}}^{\mathrm{s}}|$ and $|a_{ij, \boldsymbol{m}}^{\mathrm{t}}|$, respectively, for $\boldsymbol{m} = (1,0), (2,0), (3,0), (0,1), (1,1), (2,1)$, and $(3,1)$.
For $\sigma_i^2 \geq 4.0 \times 10^{-10}$, the coefficients that are theoretically close to zero are estimated as nonzero, reflecting noise-induced fluctuations in the phase dynamics.
The magnitude of these fluctuations is also evident in Fig.~\ref{fig:fig8}. 
Large fluctuations appear in the slow variables $\Delta \Phi$ and $\Delta \Theta$ for $\sigma_i^2 \geq 4.0 \times 10^{-10}$.
In contrast, for $\sigma_i^2 \leq 1.0 \times 10^{-10}$, each estimated coefficient agrees well with the theoretical value. 
This result indicates that the fluctuations observed in Fig.~\ref{fig:fig8} for $\sigma_i^2 \leq 1.0 \times 10^{-10}$ are sufficiently weak to allow accurate estimation of the Fourier coefficients in the phase equations.

\par
Figure~\ref{fig:fig11} shows each entry of the covariance matrix ${\rm E}_i$. 
The covariance matrix ${\rm E}_i$ is theoretically derived in Appendix~\ref{appendix:noise_derivation}.
Although the magnitudes of the estimated values are smaller than those of the theoretical ones, the estimated $E_i^{\mathrm{st}}$ and $E_i^{\mathrm{tt}}$ consistently follow the theoretically derived scaling relation $E_i^{\mathrm{pq}} \propto \sigma_i^2$ for $\mathrm{p,q} \in \{\mathrm{s}, \mathrm{t}\}$.
The scaling relation for $E_i^{\mathrm{ss}}$ breaks down in the weak-noise regime but becomes apparent as the noise intensity increases.

\section{CONCLUDING REMARKS}
\label{sec:concluding_remarks}

\par 
We developed a data-driven approach that reconstructs phase equations describing the spatiotemporal phase dynamics of reaction-diffusion systems exhibiting traveling breathers. 
We first described the data-driven approach (Sec.~\ref{sec:sec2}). We introduced the problem setting based on phase reduction theory, the calculation of the spatial and temporal phases, and the Bayesian inference process for reconstructing the phase equations. 
To validate the proposed method, we then examined whether the phase equations can be accurately reconstructed from simulation data of coupled Gray-Scott models (Sec.~\ref{sec:sec3}).

\par 
Our main results are presented in Figs.~\ref{fig:fig9} and~\ref{fig:fig10}.
Although the phase dynamics evidently fluctuate at $\sigma_i^2 = 1.0 \times 10^{-10}$ (see Fig.~\ref{fig:fig8}), they still evolve toward a linearly stable fixed point. 
In this noise regime, the deterministic part of the phase equations is accurately reconstructed.
When $\sigma_i^2 \gtrsim 4.0 \times 10^{-10}$, the phase dynamics no longer converge to the stable fixed point, and the prediction error increases to the same order of magnitude as the phase coupling functions. 
Thus, the phase equations are no longer accurately reconstructed.
The magnitude of $\sigma_i^2$ alone does not determine whether the phase dynamics can be accurately reconstructed, because the effect of noise on the phase dynamics depends on the system's sensitivity to perturbations.
Rather, successful reconstruction depends on whether the effects of phase coupling, such as convergence toward a linearly stable fixed point, can be distinguished from noise-induced phase fluctuations.
Although each entry of the covariance matrix ${\rm E}_i$ is consistently underestimated, the theoretical scaling relation with noise intensity $\sigma_i^2$ is captured (see Appendix~\ref{appendix:noise_derivation}).

\par 
From a practical viewpoint, the truncation orders of the harmonics ($M_{\mathrm{s}}$ and $M_{\mathrm{t}}$) must be reliably selected to adequately reconstruct the phase coupling functions while suppressing overfitting.
The observed spatiotemporal pattern suggests a possible criterion.
Considering $\epsilon \boldsymbol{G}_{ij}(\boldsymbol{X}_i(x,t), \boldsymbol{X}_j(x,t)) = \epsilon (\boldsymbol{X}_j(x,t) - \boldsymbol{X}_i(x,t))$, the phase coupling functions are obtained as~\cite{arai_phase_2025,kawamura_phase_2015,kawamura_phase_2019}
\begin{align}
    \label{eq:phase_coupling_function(theory)}
& \Gamma^{\mathrm{p}}_{ij}(\Phi_i-\Phi_j, \Theta_i-\Theta_j) \nonumber \\
    & = \frac{1}{2\pi}
    \int_0^{2\pi} \mathrm{d}\lambda
    \int_0^{2L} \mathrm{d}x \,
    \boldsymbol{Z}^{\mathrm{p}}(x - \Phi_i, \lambda + \Theta_i) \nonumber \\
    &\quad \cdot
    \Bigl(
    \boldsymbol{X}_0(x - \Phi_j, \lambda + \Theta_j)-\boldsymbol{X}_0(x - \Phi_i, \lambda + \Theta_i)
    \Bigr),
\end{align}
for $\mathrm{p} \in \{ \mathrm{s}, \mathrm{t} \}$.
The phase sensitivity function, $\boldsymbol{Z}^{\mathrm{p}}(x-\Phi_i, \Theta_i)$, quantifies the linear response characteristics to weak perturbations and generally reflects the spatiotemporal structure of the limit-torus solution.
The profile of the phase sensitivity function of breathers is similar to that of the limit-torus solution, reflecting the localized structure of the breather. 
That is, both spatial and temporal phases respond primarily to perturbations in the {\it core region}~\cite{arai_phase_2025}.
From Eq.~(\ref{eq:phase_coupling_function(theory)}), the phase coupling functions are determined by the structure of the limit-torus solution and the corresponding phase sensitivity functions. 
Accordingly, suitable truncation orders can be inferred from the spectral decomposition of the limit-torus solution prior to Bayesian inference.

\par
Furthermore, localized structures such as breathers provide useful information for determining these truncation orders. 
In such systems, a larger truncation order is required for the spatial phase difference than for the temporal phase difference ($M_{\mathrm{s}} > M_{\mathrm{t}}$).
This relation can be understood from the localized structure of the phase coupling functions. 
Specifically, the coupling strength remains nearly constant when the core regions of a pair of breathers remain separated but is no longer constant when the core regions overlap.
This behavior originates from the localized structure of the phase sensitivity function, which reflects the core region of the breather. 
Consistent with these findings, the phase coupling functions shown in Fig.~\ref{fig:fig4} exhibit a localized structure near $\Delta \Phi / L = 0$.

\par 
As described in Eq.~(\ref{eq:calculation_Theta}), we use linear interpolation to calculate the temporal phase, whereas alternative approaches based on signal processing, such as the method based on the Hilbert transform, are also applicable.
Appendix~\ref{appendix:Hilbert} presents the method based on the Hilbert transform for calculating the temporal phase and compares the Bayesian inference results obtained using this method with those obtained using linear interpolation.
The comparison indicates that the phase equations are more accurately reconstructed when linear interpolation is adopted than when the method based on the Hilbert transform is used.
Linear interpolation is particularly advantageous for reconstructing the temporal phase coupling function in the weak-noise regime.
Therefore, as the period of one cycle can be readily measured using the Poincar\'e section in the weak-noise regime, linear interpolation is preferable.

\par 
Our Bayesian inference formulation estimates the covariance of the noise in the spatial and temporal phases, i.e., the off-diagonal entries of ${\rm E}_i$. 
Theoretically, this covariance can be derived from the correlation between the phase sensitivity functions of the spatial and temporal phases (see Appendix~\ref{appendix:noise_derivation}). 
Although the noise covariance in the spatial and temporal phases is generally nonzero, it may be neglected in Bayesian inference when the primary interest is in accurately reconstructing the deterministic part of the phase equations. 
The problem can then be simplified by assuming all-zero off-diagonal entries. 
In Appendix~\ref{appendix:IndependentInference}, we examine whether this simplification changes the results of the Bayesian inference. 
The results remain unchanged apart from $E_i^{\mathrm{st}} = E_i^{\mathrm{ts}} = 0$.
This comparison supports the use of the simplified inference for the coupled Gray-Scott models considered here. 
However, this simplification should be applied with caution to other systems or datasets, as it may affect the accuracy of the reconstruction.

\par
Although the proposed approach reconstructs the deterministic part of the phase equations, each entry of the noise covariance matrix ${\rm E}_i$ is underestimated relative to its theoretical value.
This result suggests that the procedures used to calculate the spatial and temporal phases, as described in Sec.~\ref{subsec:CalculationOfPhase}, do not fully capture the noise-induced fluctuations expected in the corresponding asymptotic phases.
Two aspects of phase calculation may contribute to this underestimation: the method used to calculate the temporal phase and the spatial Fourier modes used to calculate the phases.
Here, we focus on the latter, whereas the former is examined separately in Appendix~\ref{appendix:Hilbert}.
Specifically, the temporal and spatial phases are calculated solely from the zero- and first-wavenumber time series, namely, $\overline{u_i}(t)$ and $\tilde{A}_i(t)$ (Eqs.~(\ref{eq:spatial_integration_u}) and (\ref{eq:tilde_A})).
In particular, the temporal phase is calculated solely from the zero-wavenumber component.
Consequently, noise-induced fluctuations present in other wavenumber components may not be fully reflected in the resulting phase time series, potentially contributing to the underestimation of $E_i^{\mathrm{pq}}$.

\par
To examine this possibility, we performed Bayesian inference using data from numerical simulations of coupled Gray-Scott models with spatially uniform noise (Appendix~\ref{appendix:SpatiallyUniformNoise}).
Under this noise setting, noise-induced fluctuations are expected to be more directly reflected in the zero-wavenumber component $\overline{u_i}(t)$ than in simulations with spatial block noise.
Because the temporal phase is calculated from $\overline{u_i}(t)$, noise-induced fluctuations in the temporal phase are expected to be captured more accurately.
Consistent with this expectation, the underestimation of $E_i^{\mathrm{tt}}$ is less severe for spatially uniform noise than for spatial block noise.

\par
Our method enables the analysis of synchronization properties of traveling and oscillating spatiotemporal patterns, whereas conventional data-driven approaches for phase reduction have primarily focused on spatiotemporal patterns that are fixed in space~\cite{fukami_data-driven_2024,yawata_phase_2025,arai_setting_2025}. 
The proposed method is expected to be useful for analyzing atmospheric circulation on Earth and fluid dynamical systems.
For example, atmospheric circulation is known to exhibit blocking phenomena, characterized by stagnation of pressure systems. 
Blocking tends to occur simultaneously in the northern and southern hemispheres~\cite{duane_synchronized_1997,duane_co-occurrence_1999}.
This synchronization can be interpreted as synchronization of amplitude variations of baroclinic waves arising from remote interactions between the two hemispheres (i.e., teleconnections). 
However, the propagation of baroclinic waves in the azimuthal direction is typically neglected. 
By incorporating this propagation of spatiotemporal patterns into the evaluation of teleconnections, our method is expected to provide new insights into large-scale climate variability.
Furthermore, our method extends the conventional analysis of teleconnections between the Arctic Oscillation and Antarctic Oscillation~\cite{tachibana_interhemispheric_2018}, in which the propagation of atmospheric circulation along the azimuthal direction is also typically neglected.
Our method is also potentially applicable to fluid systems such as rotating and oscillating convection (i.e., amplitude vacillation) in rotating fluid annuli, which can be regarded as laboratory analogs of baroclinic wave dynamics~\cite{castrejon-pita_synchronization_2010,read_phase_2017,oshima_synchronization_2025}.

\par
These atmospheric systems and rotating annulus experiments can be described in spherical and cylindrical coordinates. 
The present method is applicable to such systems, as phase reduction theory for partial differential equations with spatial translational symmetry can be formulated in these coordinate systems.
As an illustrative example, let us consider a system in which the spatiotemporal dynamics are described by a partial differential equation with a limit-torus solution.
In spherical coordinates, the spatiotemporal dynamics can be written as $\boldsymbol{X}(r, \theta, \phi, t)$, where $r$, $\theta$, and $\phi$ denote the radial, polar, and azimuthal coordinates, respectively. 
If spatial translational symmetry exists in the azimuthal direction, the limit-torus solution is given by
$\boldsymbol{X}(r, \theta, \phi, t) = \boldsymbol{X}_0(r, \theta, \phi - \Phi(t), \Theta(t))$.
Similarly, in cylindrical coordinates, the spatiotemporal dynamics can be written as $\boldsymbol{X}(r, z, \phi, t)$, where $r$, $z$, and $\phi$ denote the radial, axial, and azimuthal coordinates, respectively. 
If spatial translational symmetry exists in the azimuthal direction, the limit-torus solution is given by
$\boldsymbol{X}(r, z, \phi, t) = \boldsymbol{X}_0(r, z, \phi - \Phi(t), \Theta(t))$.
For reference, see previous studies~\cite{kawamura_phase_2015,kawamura_phase_2019} on phase reduction theory for fluid systems described by two-dimensional dynamics in a laterally periodic cylindrical geometry.
Application to experimental or observational data requires further assessment of robustness to noise, limited spatiotemporal resolution, and limited amounts of data.
Nevertheless, the present method provides a foundation for data-driven analysis of traveling and oscillating spatiotemporal patterns in systems with spatial translational symmetry.



\begin{acknowledgments}
This work was supported by JSPS KAKENHI Grant Numbers JP24H00723, JP24K06910, JP25H01474, JP25K01160, JP25K22831, JP26H00477, JP26H02349, and JP26K21344.
Numerical simulations were conducted using Earth Simulator at JAMSTEC.
\end{acknowledgments}

\section*{Data Availability}
The source code used in this study is publicly available
(\url{https://github.com/TakahiroArai-33/CODE_DATADRIVEN_RECONSTRUCTION_LIMITTORUS_PUBLIC}).

\appendix


\vspace{0.5em}  

\section{Derivation of the covariance matrix of the noise in the phase variables}
\label{appendix:noise_derivation}

\par 
In this appendix, we derive the covariance matrix ${\rm E}_i$ of the noise in the phase variables.
Here, we consider the uncoupled system and omit the index $i$.

\par 
We consider the following reaction-diffusion model with noise:
\begin{align}
    \label{eq:PDE(Appendix)}
    \frac{\partial}{\partial t} \boldsymbol{X}(x, t)
    =  \boldsymbol{F}(\boldsymbol{X}(x,t))
    + {\rm D} \frac{\partial^2}{\partial x^2} \boldsymbol{X}(x,t)
    + \boldsymbol{\xi}(x,t).
\end{align}
We assume that the reaction-diffusion system has a stable limit-torus solution representing a traveling and oscillating spatiotemporal pattern (i.e., a traveling breather) in the absence of noise (i.e., $\boldsymbol{\xi}(x,t)=\boldsymbol{0}$).
The limit-torus solution is given by
\begin{align}
    \label{eq:limit-torus(Appendix)}
    & \boldsymbol{X}(x,t) = \boldsymbol{X}_0(x - \Phi(t), \Theta(t)), ~
    \dot{\Phi}(t) = c, ~
    \dot{\Theta}(t) = \omega.
\end{align}
We assume that the noise term $\boldsymbol{\xi}(x, t)$ is Gaussian spatial block noise. In each spatial interval $x \in [\alpha \Delta, (\alpha+1)\Delta)$ with $\Delta = 2L/K$, the noise is given by
\begin{align}
    \label{eq:blocknoise_xi(Appendix)}
    \boldsymbol{\xi}(x,t) = \boldsymbol{\xi}_{\alpha}(t).
\end{align}
For $\alpha, \beta = 0, 1, \ldots, K-1$, the statistics of the spatial block noise are given by
\begin{align}
    & \langle \boldsymbol{\xi}_{\alpha}(t) \rangle 
    = \boldsymbol{0}, \\
    & \langle \boldsymbol{\xi}_{\alpha}(t)\, [\boldsymbol{\xi}_{\beta}(s)]^\mathrm{T} \rangle
    = \sigma^2\, {\rm I}\, \delta_{\alpha\beta}\, \delta(t - s).
\end{align}

\par 
On the basis of the theory for the phase reduction analysis~\cite{arai_phase_2025,kawamura_phase_2015,kawamura_phase_2019}, the phase equations are obtained as
\begin{align}
    \label{eq:PhaseEquation_beforeAverage_Phi}
    \dot{\Phi}(t)
    &=
    c
    + \sum_{\alpha=0}^{K-1}
    \boldsymbol{\zeta}_{\alpha}^\mathrm{s} \left(\Phi,\Theta \right) 
    \cdot
    \boldsymbol{\xi}_{\alpha}(t),
    \\
    \label{eq:PhaseEquation_beforeAverage_Theta}
    \dot{\Theta}(t)
    &=
    \omega
    + \sum_{\alpha=0}^{K-1}
    \boldsymbol{\zeta}_{\alpha}^\mathrm{t} \left(\Phi,\Theta \right) 
    \cdot 
    \boldsymbol{\xi}_{\alpha}(t),
\end{align}
where $\boldsymbol{\zeta}_{\alpha}^\mathrm{s} \left(\Phi,\Theta \right)$ and $\boldsymbol{\zeta}_{\alpha}^\mathrm{t} \left(\Phi,\Theta \right)$ are the effective phase sensitivity functions quantifying the linear response characteristics of the spatial and temporal phases to block noise, respectively.
The effective phase sensitivity functions are obtained as follows ($\mathrm{p} \in \{ \mathrm{s}, \mathrm{t} \}$):
\begin{align}
    \label{eq:zeta}
    \boldsymbol{\zeta}_{\alpha}^\mathrm{p} \left(\Phi,\Theta \right)\,
    &= \int_{\alpha\Delta}^{(\alpha+1)\Delta} \mathrm{d}x\, 
    \boldsymbol{Z}^{\mathrm{p}}\left(x-\Phi,\Theta \right),
\end{align} 
where $\boldsymbol{Z}^{\mathrm{s}}\left(x-\Phi,\Theta \right)$ and  $\boldsymbol{Z}^{\mathrm{t}}\left(x-\Phi,\Theta \right)$ are phase sensitivity functions quantifying the linear response characteristics of the spatial and temporal phases to weak perturbations, respectively.
The second terms on the right-hand sides of Eqs.~(\ref{eq:PhaseEquation_beforeAverage_Phi}) and (\ref{eq:PhaseEquation_beforeAverage_Theta}) are obtained as follows~\cite{masuda_collective_2010}:
\begin{align}
    & \int_{0}^{2L}
    \mathrm{d}x\;
    \boldsymbol{Z}^{\mathrm{p}}(x-\Phi,\Theta) \cdot \boldsymbol{\xi}(x,t) 
    \nonumber \\
    &=
    \sum_{\alpha=0}^{K-1}
    \int_{\alpha\Delta}^{(\alpha+1)\Delta}
    \mathrm{d}x\;
    \boldsymbol{Z}^{\mathrm{p}}(x-\Phi,\Theta) \cdot \boldsymbol{\xi}_{\alpha}(t)
    \nonumber \\
    &=
    \sum_{\alpha=0}^{K-1}
    \boldsymbol{\zeta}_{\alpha}^\mathrm{p} \left(\Phi,\Theta \right)
    \cdot \boldsymbol{\xi}_{\alpha}(t).
\end{align}

\par 
Applying the averaging method to Eqs. (\ref{eq:PhaseEquation_beforeAverage_Phi}) and (\ref{eq:PhaseEquation_beforeAverage_Theta}), the phase equations are obtained as
\begin{align}
    \label{eq:PhaseEquation_AfterAverage_Phi}
    \dot{\Phi}(t)
    &= \hat{c} + \eta^{\mathrm{s}}(t),
    \\
    \label{eq:PhaseEquation_AfterAverage_Theta}
    \dot{\Theta}(t)
    &= \hat{\omega} + \eta^{\mathrm{t}}(t).
\end{align} 
For simplicity, the noise-induced shifts are absorbed into the effective constant terms $\hat{c}$ and $\hat{\omega}$, that is,
\begin{align}
    \hat{c} &:= c+\Delta c, \\
    \hat{\omega} &:= \omega+\Delta\omega,
\end{align}
where $\Delta c$ and $\Delta\omega$ denote the $\sigma^2$-dependent noise-induced shifts of the traveling velocity and oscillation frequency, respectively.
The remaining stochastic fluctuations are represented by the zero-mean noise vector
$\boldsymbol{\eta}(t)
= (\eta^{\mathrm{s}}(t),\eta^{\mathrm{t}}(t))^{\mathrm{T}}$.
Its mean and covariance are given by
\begin{align}
    \label{eq:mean_eta(Appendix)}
    & \langle \boldsymbol{\eta}(t) \rangle = \boldsymbol{0}, \\
    \label{eq:cov_eta(Appendix)}
    & \langle \boldsymbol{\eta}(t) \left[ \boldsymbol{\eta}(s) \right]^{\mathrm{T}} \rangle
    = {\rm E} \, \delta(t-s),
\end{align}
where
\begin{align}
    \label{eq:matrix_E(Appendix)}
    {\rm E} =
    \begin{pmatrix}
    E^{\mathrm{s}\mathrm{s}} & E^{\mathrm{s}\mathrm{t}} \\
    E^{\mathrm{t}\mathrm{s}} & E^{\mathrm{t}\mathrm{t}}
    \end{pmatrix}.
\end{align}
The matrix ${\rm E}$ is symmetric, i.e., $E^{\mathrm{s}\mathrm{t}} = E^{\mathrm{t}\mathrm{s}}$.
The effective noise $\boldsymbol{\eta}(t)$ is additive; that is, its statistical properties are independent of the spatial and temporal phase variables.
Indeed, as shown below, the averaging method removes the phase dependence of the noise correlations, yielding the phase-independent covariance matrix $\mathbf{E}$.

\par 
To derive ${\rm E}$, we introduce the following correlation function:
\begin{align}
    \label{eq:correlation_C}
    C^{\mathrm{p}\mathrm{q}}(\Phi,\Theta) \nonumber 
    & = \sum_{\alpha=0}^{K-1} \sum_{\beta=0}^{K-1}
    \boldsymbol{\zeta}^{\mathrm{p}}_{\alpha}(\Phi,\Theta) 
    \cdot 
    \boldsymbol{\zeta}^{\mathrm{q}}_{\beta}(\Phi,\Theta)\, \delta_{\alpha\beta} \nonumber \\
    & = \sum_{\alpha=0}^{K-1} 
    \boldsymbol{\zeta}^{\mathrm{p}}_{\alpha}(\Phi,\Theta)
    \cdot 
    \boldsymbol{\zeta}^{\mathrm{q}}_{\alpha}(\Phi,\Theta).
\end{align}
Since the phase equations in Eqs.~(\ref{eq:PhaseEquation_AfterAverage_Phi}) and (\ref{eq:PhaseEquation_AfterAverage_Theta}) are expressed in averaged form, we consider the effective correlation of the noise.
Applying the averaging method to the correlation function, the effective correlation is obtained as
\begin{align}
    \label{eq:correlation_Cbar}
    & \overline{C^{\mathrm{p}\mathrm{q}}}
    = 
    \frac{1}{4\pi L}
    \int_{0}^{2L} \mathrm{d}x
    \int_{0}^{2\pi} \mathrm{d}\lambda \, 
    C^{\mathrm{p}\mathrm{q}}
    \left(
    x,
    \lambda
    \right). 
\end{align}
Thus, the covariance matrix ${\rm E}$ depends on the effective correlation and noise intensity as follows:
\begin{align}
    \label{eq:relation_E_and_C}
    E^{\mathrm{p}\mathrm{q}} = \sigma^2 \overline{C^{\mathrm{p}\mathrm{q}}}.
\end{align}
The scaling relation $E^{\mathrm{p}\mathrm{q}} \propto \sigma^2$ for $\mathrm{p,q} \in \{\mathrm{s}, \mathrm{t}\}$ (see the last paragraph of Sec.~\ref{subsec:Results_of_BayesianInference}) follows from Eq.~(\ref{eq:relation_E_and_C}).

\section{Calculation of the temporal phase on the basis of the Hilbert transform}
\label{appendix:Hilbert}

\par 
The temporal phase can also be calculated using signal-processing-based approaches. 
In this appendix, we describe a method based on the Hilbert transform.

\par
The protophase $\vartheta_i(t)$ is calculated from the time series of $\overline{u_i}(t)$ (Eq.~(\ref{eq:spatial_integration_u})) as follows:
\begin{align}
    \label{eq:protophase}
    \vartheta_i(t) = \arg \left[ 
        \overline{u_i}(t) + \mathrm{i}\, \overline{u_i}^{\mathcal{H}}(t)
        \right],
\end{align}
where $ \overline{u_i}^{\mathcal{H}}(t)$ is the Hilbert transform of $\overline{u_i}(t)$.
The temporal phase of an uncoupled limit-torus oscillator increases at a constant natural frequency, as described in Eq.~(\ref{eq:limit-torus}). 
However, the protophase does not exhibit such a linear time evolution and is therefore not suitable for use in the phase equation. 
To statistically rectify this problem, we transform $\vartheta_i(t)$ into $\Theta_i(t)$ as described in Refs.~\cite{kralemann_uncovering_2007,kralemann_phase_2008,kralemann_vivo_2013}, so that the resulting phase evolves linearly in time with the natural frequency.
The transformation to the temporal phase is given by
\begin{align}
    \label{eq:transformation_to_temporalphase}
    \Theta_i(\vartheta_i)
    = 2\pi \int_{0}^{\vartheta_i}  d\vartheta' \, p_i(\vartheta'),
\end{align}
where $p_i(\vartheta)$ is the probability density function of $\vartheta_i(t)$.

\par
\begin{figure*}[!t]
    \begin{center}
        \includegraphics[scale=0.95]{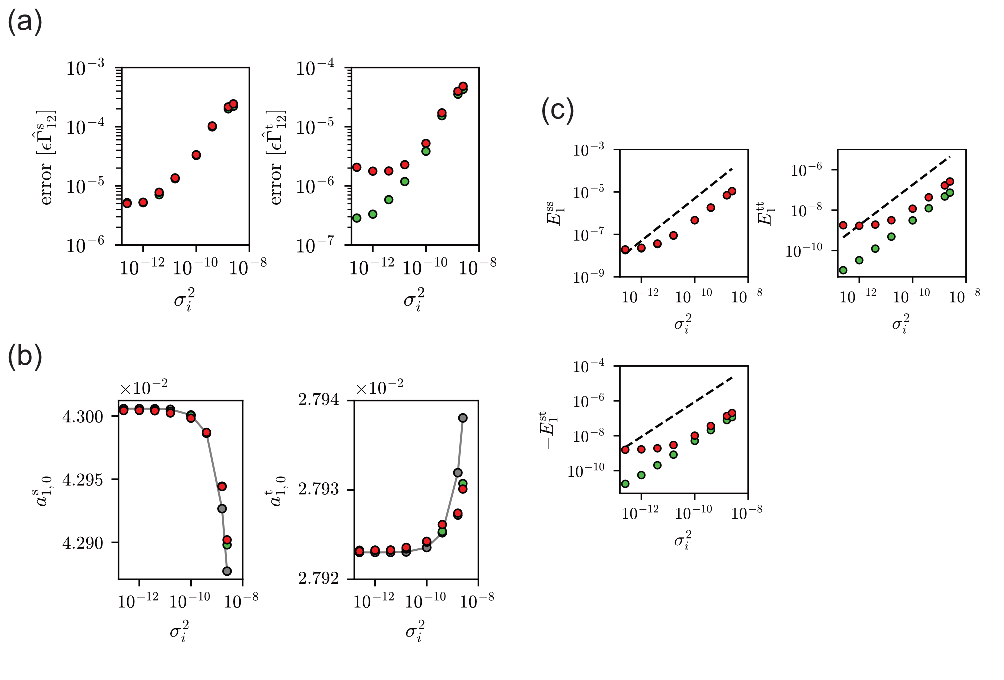}
        \caption{
        Comparison of the results of the Bayesian inference with the temporal phase calculated by the method using the Hilbert transform and the method using linear interpolation.
        Results are shown only for $(i,j) = (1,2)$.
        The red and green circles correspond to the results obtained when the temporal phase is calculated using the Hilbert transform and linear interpolation, respectively.
        (a)~
        The prediction errors of the phase coupling functions $\epsilon \hat{\Gamma}_{12}^{\mathrm{s}}$ and $\epsilon \hat{\Gamma}_{12}^{\mathrm{t}}$.
        The error is defined by Eqs.~(\ref{eq:prediction_error}) and (\ref{eq:prediction_error2}).
        Both axes are shown on logarithmic scales (log-log plot).
        (b)~
        Constant terms $a_{1, \mathbf{0}}^{\mathrm{s}}$ and $a_{1, \mathbf{0}}^{\mathrm{t}}$. 
        The gray circles indicate the reference values constructed from numerical simulations of the uncoupled Gray-Scott model~\cite{note:true_constant_terms}.
        The constant terms vary with $\sigma_i^2$ due to noise-induced shifts.
        The horizontal axis is shown on a logarithmic scale.
        (c)~
        Results of the Bayesian inference for the covariance matrix ${\rm E}_1$ versus the noise intensity $\sigma_i^2$.
        The top-left, top-right, and bottom panels show
        $E_1^{\mathrm{s}\mathrm{s}}$, $E_1^{\mathrm{t}\mathrm{t}}$, and
        $-E_1^{\mathrm{s}\mathrm{t}}$, respectively.
        The dashed lines indicate the theoretical values, which follow
        $|E_{1}^{\mathrm{pq}}| \propto \sigma_1^2$ for $\mathrm{p,q} \in \{\mathrm{s},\mathrm{t}\}$
        (see Appendix~\ref{appendix:noise_derivation}).
        Both axes are shown on logarithmic scales (log-log plot).
        }
        \label{fig:fig12}
    \end{center}
\end{figure*}

 
Figure~\ref{fig:fig12} compares the Bayesian inference results obtained using the temporal phase calculated by linear interpolation (Eq.~(\ref{eq:calculation_Theta})) and the method based on the Hilbert transform (Eqs.~(\ref{eq:protophase}) and (\ref{eq:transformation_to_temporalphase})).
Figure~\ref{fig:fig12}(a) shows the prediction errors of the phase coupling functions, Fig.~\ref{fig:fig12}(b) shows the estimated constant terms, and Fig.~\ref{fig:fig12}(c) shows each entry of the covariance matrix of the noise.
Linear interpolation yields smaller prediction errors of the phase coupling functions and constant terms closer to the true values than the method based on the Hilbert transform.
This advantage is particularly evident in the prediction errors of the temporal phase coupling function in the weak-noise regime, although $E^{\mathrm{tt}}$ and $E^{\mathrm{st}}$ are more strongly underestimated when linear interpolation is used.

\section{Simplification of the Bayesian inference with a diagonal noise covariance matrix}
\label{appendix:IndependentInference}

\par
This appendix describes a simplified Bayesian inference method with zero covariance between the noise in the spatial and temporal phases. 
We examine whether the phase equations can be accurately reconstructed under this simplification.

\par
In this simplified version of Bayesian inference, the covariance matrix of the noise is given by
\begin{align}
    \label{eq:matrix_E_cov0}
    {\rm E}_i = 
    \begin{pmatrix}
    E_i^{\mathrm{s}\mathrm{s}} & 0 \\
    0 & E_i^{\mathrm{t}\mathrm{t}}
    \end{pmatrix}.
\end{align}
In this case, we replace Eq.~(\ref{eq:update_E}) with the following update rules:
\begin{align}
    \label{eq:update_Ess_cov0}
    E_i^{\mathrm{s}\mathrm{s}}
    &=
    \frac{\Delta t}{N}
    \sum_{n=0}^{N-1}
    \bigl(
    \dot{\Phi}_{i,n}
    -
    \boldsymbol g_{i,n}^{\mathrm T}\boldsymbol a_i^{\mathrm s}
    \bigr)^{\dagger}
    \bigl(
    \dot{\Phi}_{i,n}
    -
    \boldsymbol g_{i,n}^{\mathrm T}\boldsymbol a_i^{\mathrm s}
    \bigr), \\
    \label{eq:update_Ett_cov0}
    E_i^{\mathrm{t}\mathrm{t}}
    &=
    \frac{\Delta t}{N}
    \sum_{n=0}^{N-1}
    \bigl(
    \dot{\Theta}_{i,n}
    -
    \boldsymbol g_{i,n}^{\mathrm T}\boldsymbol a_i^{\mathrm t}
    \bigr)^{\dagger}
    \bigl(
    \dot{\Theta}_{i,n}
    -
    \boldsymbol g_{i,n}^{\mathrm T}\boldsymbol a_i^{\mathrm t}
    \bigr),
\end{align}
where $\cdot^\dagger$ acting on a scalar denotes the complex conjugate. 
The remaining update rules, Eqs.~(\ref{eq:update_Xi})--(\ref{eq:update_a}), remain unchanged.

\par 
The estimation results under this simplification are consistent with those obtained without the covariance constraint; that is, the Bayesian inference results are similar to those in Figs.~\ref{fig:fig9} and \ref{fig:fig10}, apart from $E_i^{\mathrm{s}\mathrm{t}} = E_i^{\mathrm{t}\mathrm{s}} = 0$.
Therefore, this simplification does not reduce the inference accuracy for the coupled Gray-Scott models considered in this study. 
However, this simplification should be applied with caution to other systems or datasets, as it may affect the accuracy of the reconstruction.

\section{Estimation of the noise intensity $E_i^{\mathrm{tt}}$ under spatially uniform noise}
\label{appendix:SpatiallyUniformNoise}

\par
In this appendix, we apply the proposed method to simulation data from coupled Gray-Scott models with spatially uniform noise.
The results of the Bayesian inference are summarized in Fig.~\ref{fig:fig13}.

\subsection{Coupled Gray-Scott models with spatially uniform noise}

\par
For the numerical simulations, we consider a pair of coupled Gray-Scott models ($J=2$) described by Eq.~(\ref{eq:PDE_CoupledModel}) with Eqs.~(\ref{eq:GrayScottModel_X})--(\ref{eq:GrayScottModel_G}).
The parameter values are identical to those used in the numerical simulations described in Sec.~\ref{sec:sec3},
i.e., $2L=250$, ${\rm D}_i=\mathrm{diag}(1.0,1.9)$ for $i=1,2$, $\epsilon=1.0\times10^{-6}$, $f=0.018$, and $k=0.052$.
Instead of the spatial block noise defined by Eqs.~(\ref{eq:blocknoise_xi})--(\ref{eq:cov_xi}), we use spatially uniform noise defined as follows:
\begin{align}
    \label{eq:xi(InModel)_appendixD}
    \boldsymbol{\xi}_i(x,t) = \boldsymbol{\xi}_{i}(t),
\end{align}
where
\begin{align}
    \label{eq:mean_xi(InModel)}
    & \langle \boldsymbol{\xi}_{i}(t) \rangle = \boldsymbol{0}, \\
    \label{eq:cov_xi(InModel)}
    & \langle \boldsymbol{\xi}_{i}(t)\, [\boldsymbol{\xi}_{j}(s)]^\mathrm{T} \rangle
    = \sigma_i^2\, {\rm I}\, \delta_{ij}\, \delta(t - s).
\end{align}
The spatially uniform noise is expected to be more directly reflected in $\overline{u_i}(t)$, 
the time series of the zero-wavenumber component defined in Eq.~(\ref{eq:spatial_integration_u}), than in the case of spatial block noise.
Because the temporal phase $\Theta_i(t)$ is obtained from $\overline{u_i}(t)$, noise-induced fluctuations are also expected to be more clearly reflected in the time series of $\Theta_i(t)$.
We therefore examine whether the estimated noise intensity $E_i^{\mathrm{tt}}$ remains comparable to its theoretical value.

\par 
As shown in Appendix~\ref{appendix:Hilbert}, $E_i^{\mathrm{tt}}$ and $E_i^{\mathrm{st}}$ are more strongly underestimated when the temporal phase is calculated by linear interpolation than when the method based on the Hilbert transform is used.
To examine the effect of the spatial structure of the noise while avoiding the underestimation associated with linear interpolation, we therefore calculate the temporal phase using the method based on the Hilbert transform in the present analysis (Eqs.~(\ref{eq:protophase}) and (\ref{eq:transformation_to_temporalphase})).

\subsection{Theoretical derivation of the covariance matrix}
\label{subsec:uniform_noise_covariance}

\par
To provide a theoretical reference for evaluating the estimates, we derive the noise covariance matrix ${\rm E}_i$ for spatially uniform noise.
The derivation follows the procedure described in Appendix~\ref{appendix:noise_derivation}, except that the noise is assumed to be spatially uniform.
Here, we consider the uncoupled system and omit the index $i$. 
We consider the reaction-diffusion system described by Eq.~(\ref{eq:PDE(Appendix)}). 
In the absence of noise, i.e., $\boldsymbol{\xi}(x,t)=\boldsymbol{0}$, the system is assumed to have a stable limit-torus solution representing a traveling and oscillating spatiotemporal pattern. 
This solution is given by Eq.~(\ref{eq:limit-torus(Appendix)}).
We employ spatially uniform noise instead of the spatial block noise:
\begin{align}
    \label{eq:xi_AppendixD}
    \boldsymbol{\xi}(x,t) = \boldsymbol{\xi}(t),
\end{align}
where
\begin{align}
    &\langle \boldsymbol{\xi}(t) \rangle
    = \boldsymbol{0}, \\
    &\langle \boldsymbol{\xi}(t) \,
    [\boldsymbol{\xi}(s)]^{\mathrm{T}} \rangle
    = \sigma^2 {\rm I}\,\delta(t-s).
\end{align}
On the basis of the phase reduction analysis~\cite{arai_phase_2025,kawamura_phase_2015,kawamura_phase_2019}, the phase equations are obtained as
\begin{align}
    \dot{\Phi}(t)
    &=
    c
    + \int_0^{2L}\mathrm{d}x\,
    \boldsymbol{Z}^{\mathrm{s}}(x-\Phi,\Theta)
    \cdot
    \boldsymbol{\xi}(t) \nonumber \\
    &=
    c
    + \boldsymbol{\zeta}^\mathrm{s}(\Theta)
    \cdot
    \boldsymbol{\xi}(t),
    \label{eq:PhaseEquation_beforeAverage_Phi_AppendixD}
    \\
    \dot{\Theta}(t)
    &=
    \omega
    + \int_0^{2L}\mathrm{d}x\,
    \boldsymbol{Z}^{\mathrm{t}}(x-\Phi,\Theta)
    \cdot
    \boldsymbol{\xi}(t) \nonumber \\
    &=
    \omega
    + \boldsymbol{\zeta}^\mathrm{t}(\Theta)
    \cdot
    \boldsymbol{\xi}(t).
    \label{eq:PhaseEquation_beforeAverage_Theta_AppendixD}
\end{align}
The effective phase sensitivity functions, which quantify the linear response characteristics of the spatial and temporal phases to the spatially uniform noise, are obtained as follows ($\mathrm{p} \in \{ \mathrm{s}, \mathrm{t} \}$):
\begin{align}
    \label{eq:zeta_AppendixD}
    \boldsymbol{\zeta}^\mathrm{p} (\Theta )\,
    &= \int_0^{2L} \mathrm{d}x\,
    \boldsymbol{Z}^{\mathrm{p}} (x-\Phi,\Theta), 
\end{align}
where the spatial integration eliminates the dependence on $\Phi$.

\par
Applying the averaging method to Eqs.~(\ref{eq:PhaseEquation_beforeAverage_Phi_AppendixD}) and (\ref{eq:PhaseEquation_beforeAverage_Theta_AppendixD}), 
we obtain the same phase equations as Eqs.~(\ref{eq:PhaseEquation_AfterAverage_Phi}) and (\ref{eq:PhaseEquation_AfterAverage_Theta}).
The mean and covariance of the noise vector
$\boldsymbol{\eta}(t) = (\eta^{\mathrm{s}}(t),\eta^{\mathrm{t}}(t))^{\mathrm{T}}$
are given by Eqs.~(\ref{eq:mean_eta(Appendix)}) and (\ref{eq:cov_eta(Appendix)}), respectively, 
where the covariance matrix ${\rm E}$ is defined in Eq.~(\ref{eq:matrix_E(Appendix)}).
To derive ${\rm E}$, we introduce the following correlation function:
\begin{align}
    \label{eq:correlation_C_AppendixD}
    C^{\mathrm{p}\mathrm{q}}(\Theta)
    =
    \boldsymbol{\zeta}^{\mathrm{p}}(\Theta)
    \cdot
    \boldsymbol{\zeta}^{\mathrm{q}}(\Theta).
\end{align}
For the averaged phase equations, Eqs.~(\ref{eq:PhaseEquation_AfterAverage_Phi}) and (\ref{eq:PhaseEquation_AfterAverage_Theta}), the effective correlation of the noise is obtained by applying the averaging method to the correlation function:
\begin{align}
    \label{eq:correlation_Cbar_AppendixD}
    & \overline{C^{\mathrm{p}\mathrm{q}}}
    =
    \frac{1}{2\pi}
    \int_{0}^{2\pi} \mathrm{d}\lambda \,
    C^{\mathrm{p}\mathrm{q}}(\lambda).
\end{align}
Unlike in Eq.~(\ref{eq:correlation_Cbar}), the averaging in Eq.~(\ref{eq:correlation_Cbar_AppendixD}) is performed only over the temporal phase because $C^{\mathrm{p}\mathrm{q}}(\Theta)$ is independent of the spatial phase $\Phi$.
As described in Eq.~(\ref{eq:relation_E_and_C}), the entries of the covariance matrix ${\rm E}$ are determined by the effective correlation and noise intensity.
Therefore, from Eqs.~(\ref{eq:correlation_C_AppendixD}), (\ref{eq:correlation_Cbar_AppendixD}), and (\ref{eq:relation_E_and_C}), we obtain
\begin{align}
    \label{eq:E_AppendixD}
    E^{\mathrm{p}\mathrm{q}}
    &=
    \frac{\sigma^2}{2\pi}
    \int_0^{2\pi}\mathrm{d}\lambda\,
    \boldsymbol{\zeta}^{\mathrm{p}}(\lambda)
    \cdot
    \boldsymbol{\zeta}^{\mathrm{q}}(\lambda),
    \qquad
    \mathrm{p},\mathrm{q}\in\{\mathrm{s},\mathrm{t}\}.
\end{align} 
The theoretical values used for comparison with the estimated values are calculated from Eq.~(\ref{eq:E_AppendixD}).

\subsection{Results for spatially uniform noise}

\par
To obtain estimates for comparison with the theoretical values, time series of $\boldsymbol{X}_i(x,t)$ were generated through numerical simulations of the coupled Gray-Scott models.
Except for the spatial structure and intensities of the noise, all other settings, including those for time integration, spatial discretization, sampling, and the generation of the long and short time series, were the same as those used in Sec.~\ref{sec:sec3}.
The numerical simulations were performed for the following noise intensities:
$\sigma_i^2 =
2.5\times10^{-13},\,
1.0\times10^{-12},\,
4.0\times10^{-12},\,
1.6\times10^{-11},\,$
and $1.0\times10^{-10}$.

\par
\begin{figure*}[!t]
    \begin{center}
        \includegraphics[scale=0.95]{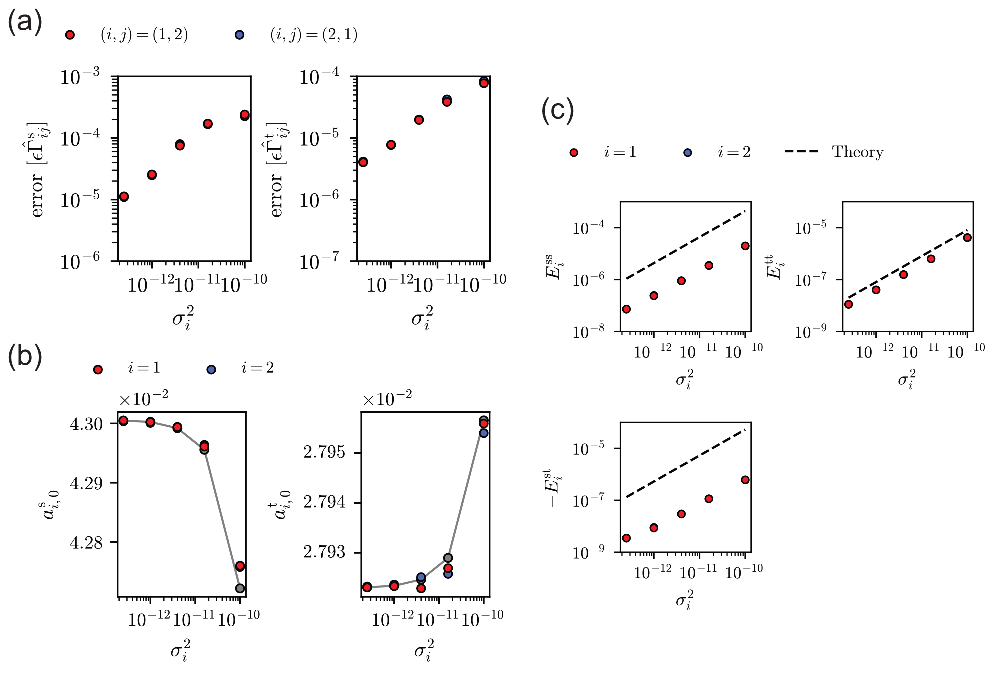}
        \caption{
            Results of Bayesian inference using numerical simulations of the coupled Gray-Scott models with spatially uniform noise.
            (a)~ 
            The prediction errors of the phase coupling functions $\epsilon \hat{\Gamma}_{ij}^{\mathrm{s}}$ and $\epsilon \hat{\Gamma}_{ij}^{\mathrm{t}}$.
            The red and blue circles correspond to $(i,j)=(1,2)$ and $(2,1)$, respectively.
            The error is defined by Eqs.~(\ref{eq:prediction_error}) and (\ref{eq:prediction_error2}).
            Both axes are shown on logarithmic scales (log-log plot).
            (b)~
            The constant terms, $a_{i, \mathbf{0}}^{\mathrm{s}}$ and $a_{i, \mathbf{0}}^{\mathrm{t}}$. 
            The red and blue circles correspond to $i=1$ and $i=2$, respectively.
            The gray circles indicate the reference values constructed from numerical simulations of the uncoupled Gray-Scott model~\cite{note:true_constant_terms}.
            The constant terms vary with $\sigma_i^2$ owing to noise-induced shifts.
            The horizontal axis is shown on a logarithmic scale.
            (c)~
            The covariance matrix ${\rm E}_i$.
            The top-left, top-right, and bottom panels show
            $E_i^{\mathrm{s}\mathrm{s}}$, $E_i^{\mathrm{t}\mathrm{t}}$, and
            $-E_i^{\mathrm{s}\mathrm{t}}$, respectively.
            The red and blue circles correspond to $i=1$ and $i=2$, respectively.
            The dashed lines indicate the theoretical values, which follow
            $|E_{i}^{\mathrm{pq}}| \propto \sigma_i^2$ for $\mathrm{p,q} \in \{\mathrm{s},\mathrm{t}\}$
            (see Appendix~\ref{appendix:SpatiallyUniformNoise}).
            Both axes are shown on logarithmic scales (log-log plot).
        }
        \label{fig:fig13}
    \end{center}
\end{figure*}

 
Figure~\ref{fig:fig13} shows the Bayesian inference results.
Figure~\ref{fig:fig13}(a) shows the prediction errors of the phase coupling functions. 
For $\sigma_i^2 \gtrsim 4.0\times10^{-12}$, the prediction errors for the spatial and temporal phase coupling functions are of order $\mathcal{O}(10^{-4})$ and $\mathcal{O}(10^{-5})$, respectively, comparable to the magnitudes of $\epsilon \hat{\Gamma}_{ij}^{\mathrm{s}}$ and $\epsilon \hat{\Gamma}_{ij}^{\mathrm{t}}$, indicating that the reconstruction is no longer reliable.
Figure~\ref{fig:fig13}(b) shows the estimated constant terms. 
The estimates of $a_{i,\boldsymbol{0}}^{\mathrm{s}}$ and $a_{i,\boldsymbol{0}}^{\mathrm{t}}$ agree well with their true values for $\sigma_i^2 \leq 1.0\times10^{-12}$. 
For $\sigma_i^2 \gtrsim 4.0\times10^{-12}$, their deviations from the true values increase, although the noise-induced shifts remain qualitatively captured.

\par
Figure~\ref{fig:fig13}(c) shows each entry of the noise covariance matrix.
Even with spatially uniform noise, $E_i^{\mathrm{ss}}$ and $E_i^{\mathrm{st}}$ remain underestimated.
By contrast, the quantity of interest, i.e., $E_i^{\mathrm{tt}}$, is substantially closer to its theoretical value.

\par 
Table~\ref{tab:Ett_noise_comparison} summarizes the theoretical and estimated values of $E_1^{\mathrm{tt}}$ shown in Figs.~\ref{fig:fig12}(c) and~\ref{fig:fig13}(c) for spatial block noise and spatially uniform noise, respectively.
Both cases use the method based on the Hilbert transform to calculate the temporal phase, but differ in the spatial structure of the noise: spatial block noise and spatially uniform noise.
We focus on the ratio of the estimated value to the corresponding theoretical value, which is also shown in the table.
For spatial block noise, this ratio generally decreases with increasing noise intensity and reaches $0.058$ at the largest noise intensity investigated.
In contrast, for spatially uniform noise, the ratio remains between $0.484$ and $0.555$ over the investigated range of noise intensities.
These results suggest that spatially uniform noise can reduce the underestimation of $E_1^{\mathrm{tt}}$, although the underestimation is not eliminated.

\par 
As derived in Appendix~\ref{appendix:noise_derivation} and Appendix~\ref{subsec:uniform_noise_covariance}, the theoretical values differ between the two noise settings even at the same input noise intensity $\sigma_i^2$.
However, conditions with comparable theoretical values can be compared, although their input noise intensities $\sigma_i^2$ differ between the two noise settings.
For example, spatial block noise with $\sigma_i^2=2.5\times10^{-9}$ and spatially uniform noise with $\sigma_i^2=1.0\times10^{-10}$ yield theoretical values of $4.48\times10^{-6}$ and $8.02\times10^{-6}$, respectively, but the corresponding ratios are $0.058$ and $0.515$.

\par
This result is consistent with the expectation that, for spatially uniform noise, noise-induced fluctuations are more directly reflected in the zero-wavenumber component $\overline{u_i}(t)$ and, consequently, in the temporal phase calculated from $\overline{u_i}(t)$.
This more direct reflection may explain why $E_i^{\mathrm{tt}}$ is less severely underestimated for spatially uniform noise.

\begin{table*}[!t]
    \caption{
        Theoretical and estimated values of $E_1^{\mathrm{tt}}$ versus the input noise intensity $\sigma_i^2$ for spatial block noise and spatially uniform noise.
        The left and right groups of columns summarize the values of $E_1^{\mathrm{tt}}$ for spatial block noise (Fig.~\ref{fig:fig12}(c)) and spatially uniform noise (Fig.~\ref{fig:fig13}(c)), respectively.
        The table also shows the ratio of the estimated value to the corresponding theoretical value.
        N/A indicates that no corresponding simulation data were used in Fig.~\ref{fig:fig13}(c).
    }
    \label{tab:Ett_noise_comparison}
    \centering
    \small
    \setlength{\tabcolsep}{4pt}
    \begin{tabular}{c @{\hspace{24pt}} r r r @{\hspace{24pt}} r r r}
        \hline\hline
        & \multicolumn{3}{c}{$E_1^{\mathrm{tt}}$ in Fig.~\ref{fig:fig12}(c)}
        & \multicolumn{3}{c}{$E_1^{\mathrm{tt}}$ in Fig.~\ref{fig:fig13}(c)} \\
        $\sigma_i^2$ & Theory & Estimate & Ratio & Theory & Estimate & Ratio \\
        \hline
        $2.5\times10^{-13}$ & $4.48\times10^{-10}$ & $1.78\times10^{-9}$ & 3.983 & $2.01\times10^{-8}$ & $1.11\times10^{-8}$ & 0.555 \\
        $1.0\times10^{-12}$ & $1.79\times10^{-9}$ & $1.69\times10^{-9}$ & 0.942 & $8.02\times10^{-8}$ & $3.99\times10^{-8}$ & 0.498 \\
        $4.0\times10^{-12}$ & $7.16\times10^{-9}$ & $1.92\times10^{-9}$ & 0.269 & $3.21\times10^{-7}$ & $1.55\times10^{-7}$ & 0.484 \\
        $1.6\times10^{-11}$ & $2.87\times10^{-8}$ & $3.16\times10^{-9}$ & 0.110 & $1.28\times10^{-6}$ & $6.26\times10^{-7}$ & 0.487 \\
        $1.0\times10^{-10}$ & $1.79\times10^{-7}$ & $1.18\times10^{-8}$ & 0.066 & $8.02\times10^{-6}$ & $4.13\times10^{-6}$ & 0.515 \\
        $4.0\times10^{-10}$ & $7.16\times10^{-7}$ & $4.29\times10^{-8}$ & 0.060 & N/A & N/A & N/A \\
        $1.6\times10^{-9}$ & $2.87\times10^{-6}$ & $1.67\times10^{-7}$ & 0.058 & N/A & N/A & N/A \\
        $2.5\times10^{-9}$ & $4.48\times10^{-6}$ & $2.60\times10^{-7}$ & 0.058 & N/A & N/A & N/A \\
        \hline\hline
    \end{tabular}
\end{table*}



\begin{thebibliography}{115}
\bibitem{note:noise_induced_shifts}
For simplicity, the noise-induced shifts of the traveling velocities and oscillation frequencies are omitted from Eqs.~(\ref{eq:PhaseEquation_Phi}) and (\ref{eq:PhaseEquation_Theta}).
These shifts are subsequently introduced as $\Delta c_i$ and $\Delta \omega_i$, respectively, and incorporated into the effective constants $\hat{c}_i$ and $\hat{\omega}_i$ (see Eqs.~(\ref{eq:define_c_hat}) and (\ref{eq:define_omega_hat})).
\bibitem{note:true_constant_terms}
To obtain the true values, we measured the traveling velocities and oscillation frequencies from numerical simulations of the uncoupled systems with noise.
We then added the theoretical values of $a_{ij,\boldsymbol{0}}^{\mathrm{s}}$ and $a_{ij,\boldsymbol{0}}^{\mathrm{t}}$ appearing in Eqs.~(\ref{eq:Gamma_s_expansion}) and (\ref{eq:Gamma_t_expansion}) to the measured traveling velocities and oscillation frequencies, respectively.
\end{thebibliography}


\end{document}